\documentclass[a4paper,11pt]{article}
\UseRawInputEncoding

\newcommand{\rh}[0]{\mathcal{R}_{H}}

\newcommand{\Dd}[0]{\mathcal{D}_{2}}

\newcommand{\hMpc}[0]{h^{-1}\mathrm{Mpc}}

\newcommand{\hGpct}[0]{h^{-3}\mathrm{Gpc}^3}

\newcommand{\LCDM}[0]{\Lambda CDM}
\newcommand{\LCDMn}[0]{$\Lambda$CDM}

\newcommand{\refEq}[1]{Eq.~\ref{#1}}
\newcommand{\refF}[1]{Fig.~\ref{#1}}

\newcommand{\refS}[1]{section~\ref{#1}}
\newcommand{\refT}[1]{table~\ref{#1}}
\newcommand{\refApp}[1]{appendix~\ref{#1}}

\pdfoutput=1 
             
\usepackage{jcappub} 

\usepackage{lineno}

\usepackage{rotating} 
\usepackage[toc,page]{appendix} 

\title{Cosmological constraints from cosmic homogeneity  \\ \today}

\author[a,1]{Pierros Ntelis,}
\author[a]{Adam James Hawken,}
\author[a]{Stephanie Escoffier,}
\author[b]{Anne Ealet,}
\author[a]{\& Andre Tilquin}

\affiliation[a]{Aix-Marseille Univ, CNRS/IN2P3, CPPM, Marseille, France}
\affiliation[b]{Institut de Physique Nucléaire de Lyon, 69622, Villeurbanne, France}
\emailAdd{pntelis@cppm.in2p3.fr}
\emailAdd{ealet@cppm.in2p3.fr}
\emailAdd{escoffier@cppm.in2p3.fr}
\emailAdd{hawken@cppm.in2p3.fr}
\emailAdd{tilquin@cppm.in2p3.fr}

\abstract{
In this paper, we study the normalised characteristic scale of transition to cosmic homogeneity, $\rh/d_V$, as a cosmological probe. We use a compilation of SDSS galaxy samples, comprising more than $10^6$ galaxies in the redshift range $0.17 \leq z \leq 2.2$ within the largest comoving volume to date, $\sim 8 \hGpct$. We show that these samples can be described by a single bias model as a function of redshift. By combining our measurements with prior Cosmic Microwave Background and Lensing information from the Planck satellite, we constrain the total matter density ratio of the universe\-$\Omega_m = 0.363 \pm 0.025$ and the Dark Energy density ratio $\Omega_{\Lambda} =  0.649 \pm 0.021$, improving the values from Planck alone by 31\% and 28\%, respectively. Our results are compatible with a flat $\Lambda$CDM model. These results show the complementarity of the normalised homogeneity scale with other cosmological probes and open new roads to cosmometry. }


\begin{document}

\maketitle
\date
\flushbottom

\section{Introduction}\label{Introduction}
The best candidate for a Standard Model of Cosmology, is known as the flat $\Lambda CDM$ model. This gives, to date, the most accurate description of our Universe mainly composed of Cold Dark Matter (CDM) and a cosmological constant $\Lambda$, responsible for the accelerating features of our cosmos.  The two main assumptions of this model are the validity of General Relativity \cite{einstein1917kosmologische} as an accurate description of gravity and the {\em Cosmological Principle} \-\cite{CP} that states that the Universe is isotropic and homogeneous on large enough scales. However, since we probe the past lightcone, we are able to test only isotropy and to link this test to homogeneity we need the Copernican Principle\cite{2011RSPTA.369.5115M}. The agreement of this model with current data is excellent, be it from type Ia supernovae\-\cite{SnIaPerlmutter,SnIaRiess,betoule2014,2018arXiv181102374D}, temperature and polarisation anisotropies in the Cosmic Microwave Background\-\cite{aghanim2018planck,CMB-Cobe}, or Large Scale Structure (LSS)\-\cite{eisenstein2005detection,LssPercival,LssParkinsonWg,LssHeymans2013Cfhtlens,BAOcosmo,2017MNRAS4702617A,2018MNRAS.473.2737S}. 

Historically, the concept of homogeneity in the large scale structure of the universe can be traced back to \citet{1760pnpm.book.....N}. As \citet{trove.nla.gov.au/work/22461641} describes, in 1932 \citet{1932AnHar..88...41S}\cite{Shapley591} published a catalogue of galaxies that was far from homogeneous, suggesting a lower limit of homogeneity of 10 Mpc, which is about the size of the Virgo Cluster. Forward in time, \citet{Martinez2} measured a characteristic homogeneity scale in the distribution of galaxies in the sky, suggesting a value larger than $100\ h^{-1}$Mpc. Since then, several methods have been developed to study the homogeneity scale\-\cite{Hogg,Yadav,Marinoni:2012ba,Sarkar,Laurent,ntelis2017exploring,2018arXiv180911125G}. Most of them found a transition to homogeneity using clustering statistics. For example in \citet{Hogg}, the authors used the fractal dimension obtained from galaxy catalogues from the Sloan Digital Sky Survey (SDSS) to estimate a characteristic homogeneity scale. Evidence for such a scale was found in other surveys as well\-\cite{Amendola, Scaramella,PanColes,kurokawa2001scaling,2012JCAP...10..036M,WiggleZ,avila2018scale,Appleby,2014arXiv1409.3831Z}. However, a debate still exists, since some authors claim not to have found such a transition and argue that the universe is non homogeneous at all scales\-\cite{,Sylos,Pietronero,1997cdc..conf...24P,Labini2,Labini3,Labini4,Joyce,park2017cosmological,2018arXiv181003539G}.

One way to estimate the scale of homogeneity is to use counts-in-spheres. 
It is expected that, in the specific case of a 3D homogeneous distribution, the number of objects inside a sphere of radius $r$ is $N(<r)\propto r^3$. While
in the more general case of a fractal distribution, $N(<r)\propto r^{D_2}$. A characteristic scale of homogeneity can then be defined as the value, $\rh$, for which the fractal dimension, $D_2$, reaches the nominal homogeneity value, $D_2=3$, to some level of precision (in our case $1\%$) for any redshift. In a recent paper, \citet{2018arXiv181009362N} (henceforth N18), we proposed using cosmic homogeneity normalised with the volume distance, as a cosmological probe to improve constraints on cosmological parameters. We provided a method to constrain the more general case of an open-$\Lambda$CDM model; using simulations that mimic the Baryon Oscillation Spectroscopy Survey (BOSS) Constant MASS (CMASS) galaxy sample\-\cite{dawson2012baryon}. In the work presented here, we extend this previous study to real data using galaxy and quasar samples from the public BOSS Data Release (DR) 12 \cite{BOSS-DR12} and extended BOSS (eBOSS) DR14 catalogues \cite{2018ApJS..235...42A}.  We use the fiducial cosmology:
\begin{equation}\label{fid-cosmo}
	p_F=(h,\omega_{b},\omega_{cdm},n_{s},\ln\left[10^{10}A_{s}  \right] ,\Omega_k) = (0.6727, 0.02225,0.1198,0.9645,3.094,0.0) \; ,
\end{equation} where $h=H_0/[100$ km s$^{-1}$ Mpc$^{-1}]$ is the dimensionless Hubble constant, $\omega_{b} = \Omega_{b}h^{2}$, is the reduced baryon density ratio, $\omega_{\rm cdm} = \Omega_{\rm cdm}h^{2}$ is the reduced cold dark matter density ratio, $n_{s}$  the spectral index, $A_{s}$ the amplitude of the primordial scalar power spectrum and $\Omega_k$ is the curvature density ratio. In this framework, the Dark Energy density ratio is defined via $\Omega_{\Lambda} = 1 - \Omega_m - \Omega_k$, where $\Omega_m = \Omega_{\mathrm{cdm}}+\Omega_{\mathrm{b}}$ is the total matter density ratio. In this analysis, we do not treat the small scales where the neutrinos have an effect in the clustering. We use 3 different fiducial cosmologies in our analysis as specified in \refApp{sec:Fiducial_cosmologies}, where one is the best cosmology as measured by Planck 2018, one is for the construction of Mock catalogues and tests in our analysis, and another one for additional tests in our analysis. We use as the default one as given by \refEq{fid-cosmo}, unless stated otherwise. 

The document is structured as follows: In \refS{sec:Dataset}, we describe the galaxy catalogues that we used in our analysis. In \refS{sec:Theory}, we describe the theoretical framework to put this analysis into context. In \refS{sec:Cosmometry}, we present the main tools that are useful in order to perform cosmometry. In \refS{sec:RH_bias_z}, we describe the bias model for the homogeneity scale. In \refS{sec:Analysis}, we describe our analysis.  In \refS{sec:Observable_Estimation}, we describe how to measure the normalised homogeneity scale in large scale structure surveys. In \refS{sec:bias_model_selection}, we describe how we select the bias model for the homogeneity scale. In \refS{sec:MCMC}, we set up the Monte Carlo Markov Chain (MCMC) analysis. In \refS{sec:Results}, we show the results of our analysis. In \refS{sec:NullTests}, we present the systematic tests that we performed to ensure the accuracy and precision of our analysis. Finally, in \refS{sec:conclusion}, we discuss our conclusions.

\section{Dataset}\label{sec:Dataset}
The Sloan Digital Sky Survey (SDSS) is a suite of surveys using the 2.5-meter Sloan Telescope\-\cite{Telescope}, located at the Apache point Observatory in New Mexico, USA. During SDSS-III\-\cite{2011AJ....142...72E}, the BOSS project\-\cite{dawson2012baryon} collected optical spectra for over a million targets. The spectroscopic galaxy sample of BOSS DR12\-\cite{BOSS-DR12} can be divided into two catalogues: the (Low Redshift) LowZ sample and the CMASS sample. 

The sky coverage of the CMASS galaxy sample is $\sim 10,200$ deg$^2$, the LowZ galaxy sample is about $\sim 9,200$ deg$^2$. Objects were selected following the CMASS and LowZ colour cuts described in \citet{TargetSelection}. 
For the CMASS sample, we selected objects in the redshift range $0.43<z<0.70$, comprising more than $800,000$ objects. For the LowZ sample, we used galaxies in the redshift range $0.172<z<0.43$, comprising of 400,000 galaxies. Note that, unlike \citet{TargetSelection}, we do not used the galaxies below $z<0.172$, since we do not have simulations at $0<z<0.172$. 

We also used the publicly available Data Release 14 \cite{2018ApJS..235...42A} of SDSS-IV from the eBOSS project \cite{dawson2012baryon},  which contains Luminous Red Galaxies (eLRG) and quasars (QSO). The eLRG sample of the extended survey covers the redshift range $0.6\leq z \leq 1.0$ over an effective area of about 2000 deg$^2$, as selected by \citet{2018ApJ...863..110B}. At higher redshifts, the QSO sample, as selected by \citet{2017Laurent} and \citet{2018Zarrouk}, covers the redshift range $0.8\leq z \leq 2.2$ with a sky coverage of 2200 deg$^2$.

\begin{table}[h!]

\caption{\label{tab:z-numbers} The number of galaxies at each redshift bin for the NGC and SGC (North and South Galactic caps) for the four different galaxy types (tracers of matter). The mean redshift at each bin is the mean between the edges of the bin. [See text for details]  }

		\begin{center} 
		\begin{tabular}{c|c|c|c} 
		  & NGC & SGC & Total \\		
		\hline
		LowZ  &  &  \\	
		\hline
		$0.172-0.258$ 	& $65194$  &  $29447$  &    \\
		$0.258-0.344$  & $97556$ &   $41723$  &    \\
		$0.344-0.430$  & $116690$  &   $48333$  &    \\
		\hline
		$0.172-0.430$  & $279440$  &   $119503$  & $398943$   \\
		\hline
		\hline
		CMASS  &  &   &    \\		
		\hline
		$0.430-0.484$ 	& $118757$  & $45601$  &    \\
		$0.484-0.538$ 	& $174385$  & $61573$  &    \\
		$0.538-0.592$ 	& $150363$  & $55336$  &    \\
		$0.592-0.700$  & $143566$  & $53531$  &    \\
		\hline
		$0.430-0.700$  & $587071$  & $216041$ & $803112$  \\
		\hline
		\hline
		eLRG  &   &   &    \\		
		\hline		
		$0.700-0.800$ 	& $20801$ & $15489$   &  $36290$   \\
		\hline
		\hline		
		QSO  &  &   &   \\		
		\hline		
		$0.800-1.150$ 	& $17911$ & $11560$  &     \\
		$1.150-1.500$ 	& $25294$ & $16743$  &     \\
		$1.500-1.850$ 	& $25614$ & $17157$  &     \\
		$1.850-2.200$  & $20422$ & $13976$  &    \\
		\hline
		$0.800-2.200$  & $89241$ & $59436$  & $148677$   \\
	 	\hline
		\end{tabular}
		\end{center}

\end{table}

Table \ref{tab:z-numbers} summarises our sample, where the number of galaxies is given for the North Galactic Cap (NGC) and the South Galactic Cap (SGC) separately. The eLRG sample was truncated to $0.7\leq z \leq 0.8$ in order to avoid correlations with CMASS and QSO samples in overlapping regions. The redshift binning was selected such as to ensure compatible statistical errors.

\subsection{Weighting scheme}
To correct for known clustering systematics, we must apply a particular weight to each galaxy. For LowZ, CMASS and eLRG  samples, we followed the weighting scheme \cite{ntelis2017exploring,TargetSelection,2018ApJ...863..110B} where we weighted each galaxy according to,
\begin{equation}
	\mathcal{W}_{gal} = w_{FKP}*w_{systot}*(w_{cp}+w_{noz} - 1) \; ,
\end{equation}
where $w_{FKP}$ is the common weight accounting for optimisations of the clustering statistics and a luminosity independent clustering bias \-\cite{2004MNRAS.347..645P}; $w_{systot}=w_{star}*w_{see}$ is the total angular systematic weight accounting for the seeing effect and the star confusion effect; $w_{cp}$ accounts for the fact that the survey cannot spectroscopically observe two objects that are closer than $62''$ and $w_{noz}$ accounts for redshift failures.
For the QSO sample, \citet{2018Zarrouk} have shown that in order to account for the efficiency of the instrument at the edges of the focal plane, and to better correct for the redshift failures, we need to treat the QSO sample with the weighting scheme:
\begin{equation}
	\mathcal{W}_{qso} = w_{FKP}*w_{systot}*w_{cp}*w_{focal} \; ,
\end{equation}
where $w_{focal}$ accounts for the inefficiency of the focal plane of the SDSS telescope. 
This weighting scheme was shown by the authors to give better estimates of anisotropic and isotropic clustering statistics. 

\subsection{Mocks, bootstraps and covariance matrices}\label{sec:mocks}
To estimate the errors and covariance matrices in this analysis we used mock catalogues and the bootstrap internal sampling method. We used the Quick Particle Mesh (QPM) mock galaxy catalogues produced by \citet{QPM}. The method is based on using quick, low-resolution particle mesh simulations that accurately reproduce the large scale dark matter density field. 
Particles are then sampled from the density field based on their local density such that they have N-point statistics nearly equivalent to the halos resolved in
high-resolution simulations. These simulations are used to create a set of mock halos that can be populated using halo
occupation methods to create galaxy mocks. Then the survey geometry is imprinted on those catalogues to produce the mock catalogues that we use in this study. 
The cosmology used to obtain these catalogues is:
\begin{equation}\label{qpm-cosmo}
	p_{qpm}=(h,\omega_{b},\omega_{cdm},n_{s},\ln\left[10^{10}A_{s}  \right] ,\Omega_k) = (0.7, 0.02247,0.1196,0.97,3.077,0.0) \; .
\end{equation}
The SDSS collaboration has made the QPM mock catalogues for the  LowZ and CMASS samples publicly available. We used $100$ of them to compute the covariance matrix of the fractal dimension $\Dd{}$, see \refS{sec:Theory}.

Mock catalogues were not available at the time for the  eLRG and QSO samples. To compute covariance matrices and perform tests for the eLRG and QSO samples, we used a bootstrap internal sampling method. The bootstrap method consists of subsampling each galaxy catalogue with replacement. 

Then we computed $\Dd{}$ for each sub-sampled catalogue to produce the covariance matrix. For either method the covariance matrix is given by:
\begin{equation}\label{eq:Covariance_Fractal}
	C_{ij} = \frac{1}{N_r-1} \sum_{n=1}^{N_r} \left( \mathcal{D}^{(n)}_2(r_i) - \bar{\mathcal{D}}_2(r_i) \right) \left( \mathcal{D}^{(n)}_2(r_j) - \bar{\mathcal{D}}_2(r_j) \right) \; ,
\end{equation}
where $N_r$ is the number of realisations. For our fitting method, we corrected our precision matrix following \citet{2013MNRAS.432.1928T}, using $\psi_{ij} =\frac{N_r-N_d-2}{N_r-1} C^{-1}_{ij}$, where $N_d$ is the number of data bins. In \refApp{sec:Bootstrap_covariance_validation}, we quantify the validity of the bootstrap method. 

\section{Method}\label{sec:Theory}


In this paper, firstly, we are interested in the theoretical prediction of a characteristic scale of the homogeneity scale of the universe normalised with the volume distance, $\rh/d_V$, for a given theoretical model, in our case the open-$\Lambda$CDM model. The homogeneity scale, following N18 and references therein, can be defined as the scale at which the fractal dimension, $\Dd{}$, takes the value corresponding to a three dimensional homogeneous distribution to within $1\%$ precision, formally written as:
\begin{equation}\label{eq:RH-definition}
 \mathcal{D}_{2}(\rh) \equiv 2.97 \; ,
\end{equation}
where the fractal dimension is given by,
\begin{equation}\label{eq:d2observable}
	\mathcal{D}_2(r) = \frac{d\ln }{d\ln r} \left[  1+ \frac{3}{r^{3}} \int_{0}^{r}{\xi}(s)s^{2}ds  \right] + 3 \; ,
\end{equation}
where $\xi$ is the two-point correlation function. The two point correlation function is related to the Power Spectrum, $P(k)$, through the Fourier Transform,
\begin{equation}\label{eq:real_space_correlation_function}
	\xi (r) = \frac{1}{2\pi^2} \int dk k^{2} \frac{\mathrm{sin}(kr)}{kr} P(k) \; .
\end{equation}

Equation \ref{eq:real_space_correlation_function} describes the theoretical prediction of the correlation function of the total matter of the universe in real space, quantified by $P(k)$ or $\xi(r)$. However, we measure the correlation function of luminous galaxies in redshift space. The distribution of luminous galaxies is biased with respect to that of the total matter of the universe. Thus we must include a model for this effect. Furthermore, the redshift of each galaxy has contributions due the peculiar motions of that galaxy. These contributions induce distortions in the clustering of galaxies in redshift space. Therefore we must also include a model for these redshift space distortions.
\citet{kaiser} and \citet{hamilton1992measuring} have shown, that on our scales of interest, the monopole of the two-point correlation function in redshift space for a biased tracer, with bias, $b(z)$, is given by: 
\begin{equation}\label{eq:kaiser_model}
	\xi_0^{(s,G)}(r,z) = \frac{1}{2}\int^{1}_{-1}d\mu\ \xi^{(r)}(r,\mu,z) = b^{2}(z)\left[ 1 + \frac{2}{3}\frac{f(z)}{b(z)} + \frac{1}{5}\left( \frac{f(z)}{b(z)} \right)^2 \right] \xi^{(r,m)}_0(r,z) \; ,
\end{equation}
where the superstcript $(r)$ denotes real space; $(s)$ denotes redshift space; $(G)$ denotes the galaxy tracer; $(m)$ denotes the total matter of the universe ( We give more details on the $b(z)$ model in \refS{sec:RH_bias_z} ); the $f(z)$ is the usual growth factor, which is modelled as:
\begin{equation}
	f(z) \simeq \Omega_m^{\gamma}(z) =  \left[  \frac{  \Omega_m (1+z)^3 }{ E^2(z;\Omega_m,\Omega_{\Lambda}) }\right]^{\gamma} 
\end{equation}
where in the case of standard Einstein's Gravity, $\gamma = 0.545$, as shown by\-\citet{Growth}. Note that:
\begin{equation}\label{eq:Hz}
	E(z) \equiv \frac{H(z)}{H_{0}} =  \sqrt{ (\Omega_{\rm cdm} +\Omega_{\rm b} )(1+z)^{3} + \Omega_{\Lambda}  +\Omega_k(1+z)^2 } \; ,
\end{equation}
is the usual normalised Hubble expansion rate as a function of redshift. Substituting \refEq{eq:kaiser_model} into \refEq{eq:d2observable}, we get the biased redshift space distorted fractal dimension, 
\begin{equation}\label{eq:d2observable_galaxies}
	\Dd{}^{G}(r;z,b_0,\Omega_m,\Omega_{\Lambda}) =  3 + \frac{d\ln}{d\ln\ r} \left[ 1 + \frac{3}{r^3} \int^{r}_{0} ds s^2 \xi_0^{(s,G)}(s,z;b_0,\Omega_m,\Omega_{\Lambda}) \right]
\end{equation}
Our homogeneity threshold (Omitting the parameter dependence for clarity) is then: 
\begin{equation}\label{eq:RH-definition_galaxies}
	\Dd{}^{G}(\rh^{G}) \equiv 2.97 \; ,
\end{equation}
where $\rh^{G}$ is explicitly defined as:
\begin{equation}\label{eq:rh_in_threshold}
	\rh^{G} \equiv \rh^{G}(z;b_0,\Omega_m,\Omega_{\Lambda}) \; ,
\end{equation}
where we have explicitly restored the parameter dependence. We can also consider other parameters that $\rh$ depends upon, such as the neutrino masses, but they are not relevant to the scales that we are probing.

It is convenient to introduce the cube of the \textit{volume distance} (or comoving volume element), 
\begin{equation}\label{eq:d_V}
	d_V^3(z) = \frac{cz}{H(z)} d_M^2(z) \; ,
\end{equation}
where $d_M$ is the \textit{motion distance} (or transverse comoving distance) and $c$ is the speed of light.

Now, following \citet{2015A&A...584A..69R}, we normalised the homogeneity scale using  \refEq{eq:d_V}, which defines the model of our observable, (in other words, the theoretical predictions):
\begin{equation}\label{eq:model_y}
	M(z;b_0,\Omega_m,\Omega_{\Lambda}) = \frac{\rh^G(z;b_0,\Omega_m,\Omega_{\Lambda})}{d_V(z;\Omega_m,\Omega_{\Lambda})}
\end{equation}
for different redshift slices, for a given bias model and cosmology. This normalisation ensures that the observable is independent of the $h$ parameter. The division with the volume distance, takes into account the isotropic dilations of the homogeneity scale at different redshifts. 

 We note that even though the homogeneity scale is not an one-to-one function of the $\Omega_m$ parameter for a flat-$\Lambda$CDM cosmology, the normalised homogeneity scale is very sensitive to both $(\Omega_m,\Omega_{\Lambda})$ parameters for an open-$\Lambda$CDM cosmology, as we show in \refS{sec:Sensitivity_of_the_normalised_homogeneity_scale}. 

Keep in mind that our observable is biased with respect to the total matter of the universe, therefore we take that into account as we explain in \refS{sec:RH_bias_z}. Additionally, the likelihood may be biased towards the fiducial cosmology and we study this effect in \refS{sec:NullTests}.

We study the cosmological parameter space $p=(b_0,\Omega_m,\Omega_{\Lambda})$. Note that $\Omega_m,\Omega_{\Lambda}$ corresponded to the observed parameters, the ones that we measure, which are different from the $\Omega^F_{m},\Omega^F_{\Lambda}$ given from previous knowledge. We note the rest of the parameters remained unchanged in the framework of the fiducial cosmology that we used for each procedure. We note that $A_s$ parameters is degenerate with the $b_0$ since it changes only the amplitude of the Power Spectrum as we explained in N18. We also note that one can compute the $\sigma_8$ parameter given the computed Power Spectrum, the value of the $A_s$ and the Window Function of the survey. Since we fix $A_s$, we do not consider constrains in the $\sigma_8$ parameter in this study. 

In order to retrieve all these related quantities, 
we used the CLASS code \cite{CLASS} to solve the perturbed Einstein-Boltzman equation to obtain $P(k)$ for a given cosmology. Therefore we compute iteratively for the chosen to study parameters $p=(b_0,\Omega_m,\Omega_{\Lambda})$ the \refEq{eq:d2observable}, \refEq{eq:rh_in_threshold} and \refEq{eq:model_y} and we keep fixed the rest parameters to their fiducial values. For example, we use 1 massive and 2 relativistic neutrinos, with $N_{eff} = 3.046$. We have made our codes for the computation of the homogeneity scale and related quantities publicly available\footnote{\url{https://github.com/lontelis/cosmopit} }.

\subsection{Cosmometry with $\rh/d_V$}\label{sec:Cosmometry}
From the observational point of view, we need to infer distances from ($z,R.A.,Dec$) positions of galaxies. Therefore, we transform them into comoving coordinates using the \textit{comoving distance}:
\begin{equation}\label{eq:FRW}
		d_{\rm C} (z) = d_H \int_{0}^{z} dz'E^{-1}(z') \;,
	\end{equation}
where $d_H=c/H_0$ is the hubble distance today and $c$ is the speed of light. 
Having these tools in hand, assuming a flat-$\LCDM$ model (i.e. using the fiducial cosmology given by \refEq{fid-cosmo}), as well as equations \ref{eq:FRW} and \ref{eq:Hz}, we transform the redshift of each galaxy to a radial comoving distance. This gives the comoving positions of the galaxies in three dimensional redshift space. 

Now, we use the \citet{LSestimator} estimator to extract the monopole of the two-point correlation function from the positions of galaxies, using the CUTE software \cite{alonso2012cute}. From this estimate, using \refEq{eq:d2observable_galaxies} and \refEq{eq:RH-definition_galaxies}, we extract the fractal dimension, $\mathcal{D}_2^G(r)$, and the homogeneity scale, $\rh^G$ in each redshift slice.  Now, using \refEq{eq:d_V}, we normalise the data, which result to our observable:
\begin{equation}\label{eq:observable_real_space}
		O(z;\Omega_m^{F},\Omega_{\Lambda}^{F}) = \frac{\rh^G(z;\Omega_m^F,\Omega_{\Lambda}^F)}{d_V(z;\Omega_m^F,\Omega_{\Lambda}^F)}
\end{equation}
for different redshift slices and for a given fiducial cosmology denoted by the parameters $(\Omega_{m}^F, \Omega_{\Lambda}^F$).

\subsection{Cosmic bias model for $\rh/d_V$}\label{sec:RH_bias_z}

In this section, we explain the construction of the single parameter bias model for the characteristic normalised homogeneity scale and in \refS{sec:bias_model_selection} we justify its use.

The homogeneity scale for a given tracer of matter, such as the galaxy distribution, $\rh^G$, is related to the homogeneity scale of the matter distribution, $\rh^m$, up to a redshift space distortion model and a bias model, $b(z)$ as discussed in \refS{sec:Theory} through the definition of equations \ref{eq:RH-definition}, \ref{eq:d2observable} and \ref{eq:kaiser_model}. We construct the bias model as follows.

The different tracers we are studying here (LowZ, CMASS, eLRG and QSO) emit light via different physical processes. For example, CMASS galaxies are passively evolving massive galaxies whose emission is composed of star light \cite{Maraston}, on the other hand, QSOs are active galaxies whose emission is caused by accretion around a central super massive black hole \cite{2012ApJS..199....3R}. Thus, the nature of the relationship between the distributions of luminous and dark matter traced by these objects will be different. There are, therefore, physically motivated reasons to model the bias differently for each sample. In \refS{sec:bias_model_selection}, we measure the biases by applying the bias model individually at each different sample.

However, implementing multiple bias models would require introducing a large number of bias parameters. Having to marginalise over all these parameters would not be possible with this data set. Therefore, we investigate the efficacy of two different single effective bias parameters.

The first one is:
\begin{equation}
	b_{1,\rh}(z) = b_0\sqrt{1+z}  \; ,
\end{equation}
following \citet{2018LRR....21....2A,montanari2015measuring}.

The second one is a piecewise linear bias model as 
a function of redshift. 
For lower redshifts $z<z_{\star}$, we use the linear bias:
\begin{equation} \label{eq:bias_model1}
	b^{z<z_{\star}}(z;b_0) = b_0\sqrt{1+z} \; ,
\end{equation}
while for the higher redshift QSO sample, the cosmic bias, according to \citet{2017JCAP...07..017L}, is:
\begin{equation}\label{eq:bias_model_qso}
	b^{z>z_{\star}}(z;b_1,b_2) = b_1*(1+z)^2 + b_2 \; .
\end{equation}
We make the assumption that there is a continuity between the lower redshifts and higher redshifts at $z\sim z_{\star}$ and therefore we impose the following continuity conditions between the two redshift regions:
\begin{align}
	b^{z<z_{\star}}(z=z_{\star} ) &= b^{z>z_{\star}}(z=z_{\star} )  \label{eq:bias_condition_1}\\
	\frac{ \partial b^{z<z_{\star}}}{\partial z}  |_{z=z_{\star} } &= \frac{ \partial b^{z>z_{\star}}}{\partial z}  |_{z=z_{\star} } \label{eq:bias_condition_2}
\end{align}
After some algebra we find that:
\begin{align}
	b_1 &= \frac{1}{4} \frac{1}{ (1+z_{\star})^{3/2} } b_0 \; , \\ 
	b_2 &= \frac{3}{4}\sqrt{1+z_{\star}} b_0 \; . 
\end{align}
Now, \refEq{eq:bias_model_qso} can re-written, as a function of $b_0$ and $z_{\star}$ as:
\begin{equation}
	b^{z>z_{\star}}(z;b_0) = b_0 \left[   \frac{1}{4} \frac{1}{ (1+z_{\star})^{3/2} } (1+z)^2 + \frac{3}{4}\sqrt{1+z_{\star}} \right] \label{eq:bias_z_qso}\; .
\end{equation}
Therefore, the second parametrisation of the cosmic bias model for the homogeneity scale at redshifts $0.0 < z < 2.2$ can be written as :
\begin{equation}\label{eq:bias_model2}
		b_{2,\rh}(z;b_0,z_{\star})  = b_0
		\left\{ 
			\begin{matrix}
				\sqrt{1+z} \hspace{4.4cm},\ \mathrm{for}\ z<z_{\star}  \\
				\left[   \frac{1}{4} \frac{1}{ (1+z_{\star})^{3/2} } (1+z)^2 + \frac{3}{4}\sqrt{1+z_{\star}} \right]\ ,\ \mathrm{for}\ z>z_{\star} 
			\end{matrix} 
		\right\} 
		\; . 
\end{equation}
We choose $z_{\star}=0.8$ which is the redshift where the QSO sample starts. In \refS{sec:bias_model_selection}, we test both of the above bias models against the data.

\section{Analysis}\label{sec:Analysis}
In this section, we describe our analysis given the method described in \refS{sec:Theory}. We briefly describe the estimation of the normalised homogeneity scale as obtained from the different galaxy catalogues. Then we describe the method that we used in order to choose the best bias model for the homogeneity scale. Then we present the Monte Carlo Markov Chain (MCMC) analysis that we performed in order to constrain cosmology. We also present the test against the fiducial cosmology. 

\subsection{Observable estimation}\label{sec:Observable_Estimation}

For each of the galaxy samples described in \refS{sec:Dataset}, the fractal dimension, $\mathcal{D}^G_2(r)$, as defined in \refEq{eq:d2observable_galaxies}, is computed over the range $r=[50,200] \hMpc$, in each redshift bin. We used the 100 QPM mocks (or 100 bootstraps, see \refS{sec:mocks}) for the different redshift bins to construct the covariance matrix of $\Dd{}^G$. We tested the validity of using bootstraps in the cases where we do not have mocks in \refApp{sec:Bootstrap_covariance_validation}. The function, $\Dd{}^G$, is then fitted by a spline. The homogeneity scale of the galaxy samples, $\rh^G$ is then the scale at which this spline crosses $\Dd{}^G=2.97$,
extracted using the definition given in \refEq{eq:RH-definition_galaxies}.  

We have made measurements in different redshift bins and in two different fields, the North Galactic Cap and the South Galactic Cap. We wanted to know whether the different redshift bins are independent. Therefore, we studied the correlation coefficient, defined as $\rho = C_{z_iz_j}/\sqrt{C_{z_iz_i}C_{z_j z_j}}$ of $\rh/d_V$, between redshift bins using the mock and bootstrap catalogues.  This means that for the North Galactic Cap (South) we can define the covariance matrix:
\begin{equation}\label{eq:Covariance_observable}
	C_N \equiv C_{z_iz_j} = \frac{1}{N_r-1} \sum_{n=1}^{N_z} \left( O^{(n)}(z_i) - \bar{O}(z_i) \right) \left( O^{(n)}(z_j) - \bar{O}(z_j) \right)\; ,
\end{equation}
where $O$ is given by \refEq{eq:observable_real_space}, $N_r$ is the number of realisation, $N_r=100$. The superscrit $N$ denotes the North galactic cap, we use $S$ for the South. We found the correlation coefficient to be $|\rho|\lesssim0.25$ for the NGC and $|\rho|\lesssim 0.20$ for the SGC, as shown in \refApp{sec:Correlation_matrix_rhdv}. These values show that the covariance between redshift bins is non-negligible. Furthermore, we need to combine measurements in the two fields into one number for each redshift. Therefore, we define the weighted average of the normalised homogeneity scale as:

\begin{equation}\label{eq:data}
	\frac{\rh^G}{d_V} (z) = \left( C_{N}^{-1} + C_{S}^{-1} \right)^{-1} \left( C_{N}^{-1}\frac{\rh^{N}}{d_V}(z) + C_{S}^{-1}\frac{\rh^{S}}{d_V}(z) \right) \; ,
\end{equation}
where the superscript N denotes the NGC and S the SGC. We also combine the covariance matrix in the usual way:
\begin{equation}\label{eq:covmatrix}
	C_{\rh/d_V} (z_i,z_j) = \left( C_{N}^{-1} + C_{S}^{-1} \right)^{-1} \; .
\end{equation}

Table \ref{tab:RH-z} shows the estimated homogeneity scale, $\rh^G$, the theoretical homogeneity scale for the total matter without redshift space distortions, $\rh^m$, and the volume distance, $d_V$, for the different galaxy samples in the different redshift bins. 


\begin{table}[h!]

\caption{\label{tab:RH-z} Mean and standard deviation of the homogeneity scale, $\rh$, as a function of redshift, $z$, for the different galaxy samples as explained in \refS{sec:Observable_Estimation}. The second to last column is the expected homogeneity scale for the total matter of the universe, $\mathcal{R}^m_H$, while the last column is the fiducial volume distance, see $d_V$ \refS{sec:Observable_Estimation}. }

		\begin{tabular}{c|c|c|c|c|c} 
		 Type & $\Delta z$ & $z_{eff}$ & $\mathcal{R}^G_{H} [h^{-1}$Mpc] & $\rh^m [h^{-1}\mathrm{Mpc}]$ & $d_V [h^{-1}\mathrm{Mpc}]$\\
		\hline 
		\hline
		LowZ &$0.172-0.258$ & $0.215$ & $109.89\pm8.76$  &   73.87 & $600$   \\
			  &$0.258-0.344$ & $0.301$  & $136.49\pm7.54$  &    70.87 & $813$  \\
		          &$0.344-0.430$ & $0.387$  & $126.69\pm6.85$ &    68.0 & $1012$   \\
		\hline
		\hline
		CMASS    &$0.430-0.484$ & $0.457$ 	&  $121.55 \pm6.24$  &   65.80 & $1163$ \\ 
				&$0.484-0.538$ & $0.511$ 	&  $117.00 \pm 4.94$  &   64.16 & $1274$ \\
				&$0.538-0.592$ & $0.565$ 	&  $116.57 \pm 4.83$  &   62.55 & $1379$ \\
				&$0.592-0.700$ & $0.646$  &  $118.79 \pm 3.19$  &   60.25 & $1529$ \\
		\hline
		\hline		
		eLRG    &$0.700-0.800$ & $0.750$ 	&  $117.12 \pm 5.38$ &  $57.45$ & $1704$   \\	
		\hline
		\hline
		QSO	 &$0.800-1.150$  & $0.975$	&  $95.16 \pm 8.37$ & $51.96$ & $2033$ \\ 
			 &$1.150-1.500$ & $1.325$ 	&  $97.90 \pm 6.48$ & $44.78$ & $2430$ \\
			 &$1.500-1.850$ & $1.675$ 	&  $96.44 \pm 5.42$ & $38.96$ & $2726$\\ 
			 &$1.850-2.200$ & $2.025$       &  $105.23 \pm 9.50$ &  $34.17$  & $2952$\\ 
		\end{tabular}
\end{table}

\subsection{Bias model selection}\label{sec:bias_model_selection}

There are different bias models in the literature. \cite{2018LRR....21....2A,montanari2015measuring} have shown that the two-point correlation function at low redshifts can be modelled using a linear bias model, which is given by \refEq{eq:bias_model1}. At high redshifts, $z\gtrsim0.8$, \citet{2017JCAP...07..017L}, and reference therein, modelled the bias using \refEq{eq:bias_model_qso}. In order to investigate whether at all redshifts we can use a single bias model, we performed the following test. 

We fitted a bias model (selected based on the redshift of the sample) to each sample separately, keeping the cosmology fixed to our fiducial cosmology. To the LowZ, CMASS and eLRG samples, we fitted the low redshift bias model described by \refEq{eq:bias_model1}, which has a single bias paramter, $b_0$. While for the higher redshift, QSO sample, we applied the high redshift bias model described in  \refEq{eq:bias_model_qso} which has two free parameters, $b_1,b_2$.


	   \begin{figure}[h!]
	   \hspace{-1.0cm} 
	    \includegraphics[width=180mm]{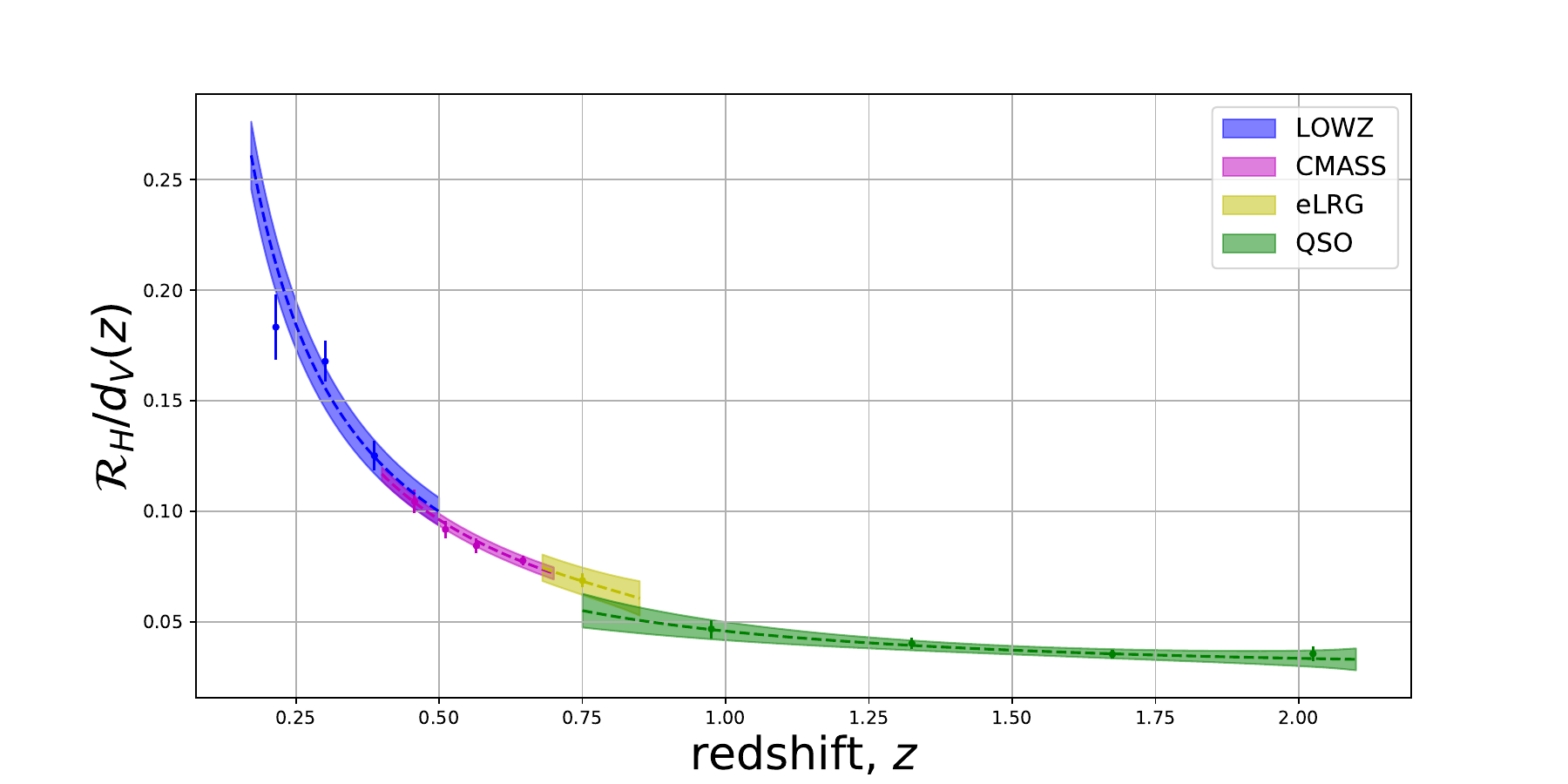}
	    	    \caption{\label{fig:bias_model_testValidation} The normalised homogeneity scale as a function of redshift, for the different galaxy samples, colorcoded. The shade areas show the $2\sigma$ deviation of the fiducial model fitted to the data for the different parametrisation of the bias for the different galaxies sample. [see \refS{sec:bias_model_selection}] }
	   \end{figure}%

Figure \ref{fig:bias_model_testValidation} shows the normalised homogeneity scale as measured in our four galaxy samples, LowZ (blue), CMASS (purple), eLRG (yellow) and QSO (green). The coloured bands represent the $2\sigma$ uncertainty on the best fitting bias parameters for each sample, as described above. These models have been extrapolated beyond the redshift range of the corresponding data. The overlap between the bands shows that the models are compatible with one another. For example, the best fitting bias model for the QSO sample and the best fitting bias model for the eLRG sample, are compatible to the level of $1.4\sigma$. 
We use this as justification for adopting the same bias model at all redshifts. 

The result for one fiducial cosmology, $p_F$ is shown in \refF{fig:bias_model_testValidation}. We repeated this test using a $p_{F2}$ fiducial cosmology, which is defined by:
\begin{equation}\label{eq:fid-cosmology-test}
		p_{F2}=(h,\omega_{b},\omega_{cdm},n_{s},\ln\left[10^{10}A_{s}  \right] ,\Omega_k) = (0.7, 0.0225,0.11172,0.95,3.077,0.0) \; ,
\end{equation}
on the data and we obtained similar results.

In \refS{sec:RH_bias_z}, we have described two single parameter bias models for all redshift bins. 
To select one of the two candidates, \refEq{eq:bias_model1} and \refEq{eq:bias_model2}, we investigated their $\chi^2$. In \refT{tab:Tests_bias}, we show that the first model performs better since its $\chi^2$ is closer to the number of degrees of freedom. We repeated this test using a $p_{F2}$ fiducial cosmology and we obtained similar results. Henceforth, all the results we present here, have been obtained with the first bias model.

\begin{table}[h!]

  \caption{\label{tab:Tests_bias} Measurement of the cosmological parameters for the two different bias models, as described in \refS{sec:bias_model_selection}, considering the combination, $b_{i,\mathcal{R}_H}$-$\mathcal{R}_H/d_V$$+$CMB$+$Lensing. 
  }
  
\vspace{1mm}
\center
\begin{tabular}{l | c c c c} \hline
 Bias model & $b_0$ & $\Omega_m$ & $\Omega_{\Lambda}$ & $\chi^2\pm\sqrt{2.ndf}$, $ndf=11$ \\ 
 \hline
 

    
     \hline
$b_{1,\mathcal{R}_H}$ &

 $1.397 \pm 0.026$ &
 $0.363 \pm 0.025$ &
 $0.650 \pm 0.021$ &
 $7.524 \pm 4.690$
 \vspace*{0.0mm} \\
$b_{2,\mathcal{R}_H}$ &

 $1.397 \pm 0.026$ &
 $0.360 \pm 0.025$ &
 $0.652 \pm 0.020$ &
 $7.397 \pm 4.690$
 \vspace*{0.0mm} \\
 \hline 
 \end{tabular}
 
 \end{table}


\subsection{Sensitivity of the normalised homogeneity scale to cosmology}\label{sec:Sensitivity_of_the_normalised_homogeneity_scale}

\citet{2019arXiv190102400N}  have argued that the homogeneity scale cannot be used as a cosmological probe. In particular, they have shown that the homogeneity scale does not have a one-to-one relationship with the $\Omega_m$ parameter. However, in this paper as in N18, we are studying the homogeneity scale normalised with the volume distance, $\rh/d_V$. We stress that even though the $\rh$ alone is not one-to-one function with $\Omega_m$ parameter for a flat-$\Lambda$CDM model the parameter combination, $\rh/d_V$, does have a cosmological dependence, as we show below. 

	   \begin{figure}[h!]
	   \hspace{-0.0cm} 
	    \includegraphics[width=0.49\linewidth, keepaspectratio]{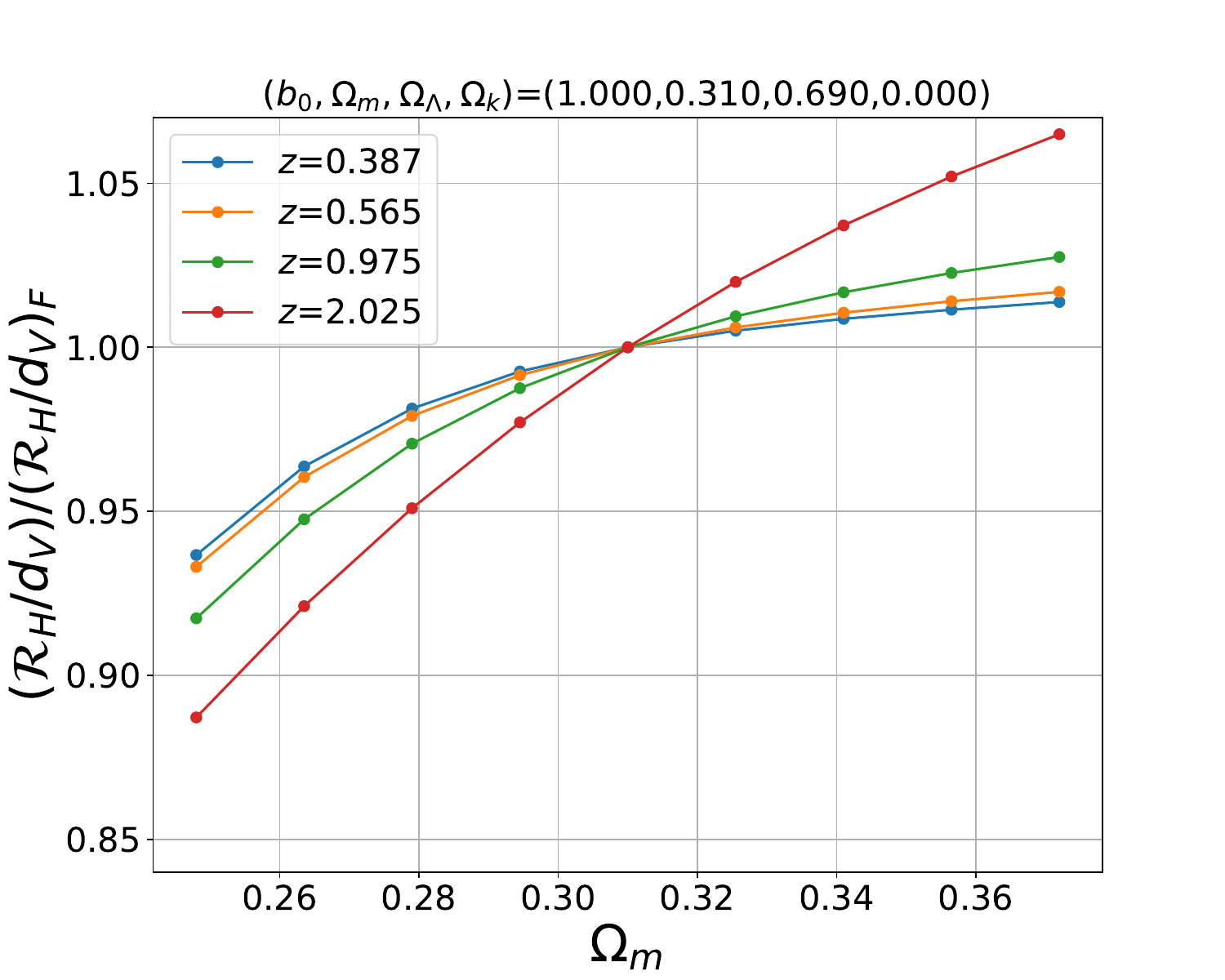}
	    \includegraphics[width=0.49\linewidth, keepaspectratio]{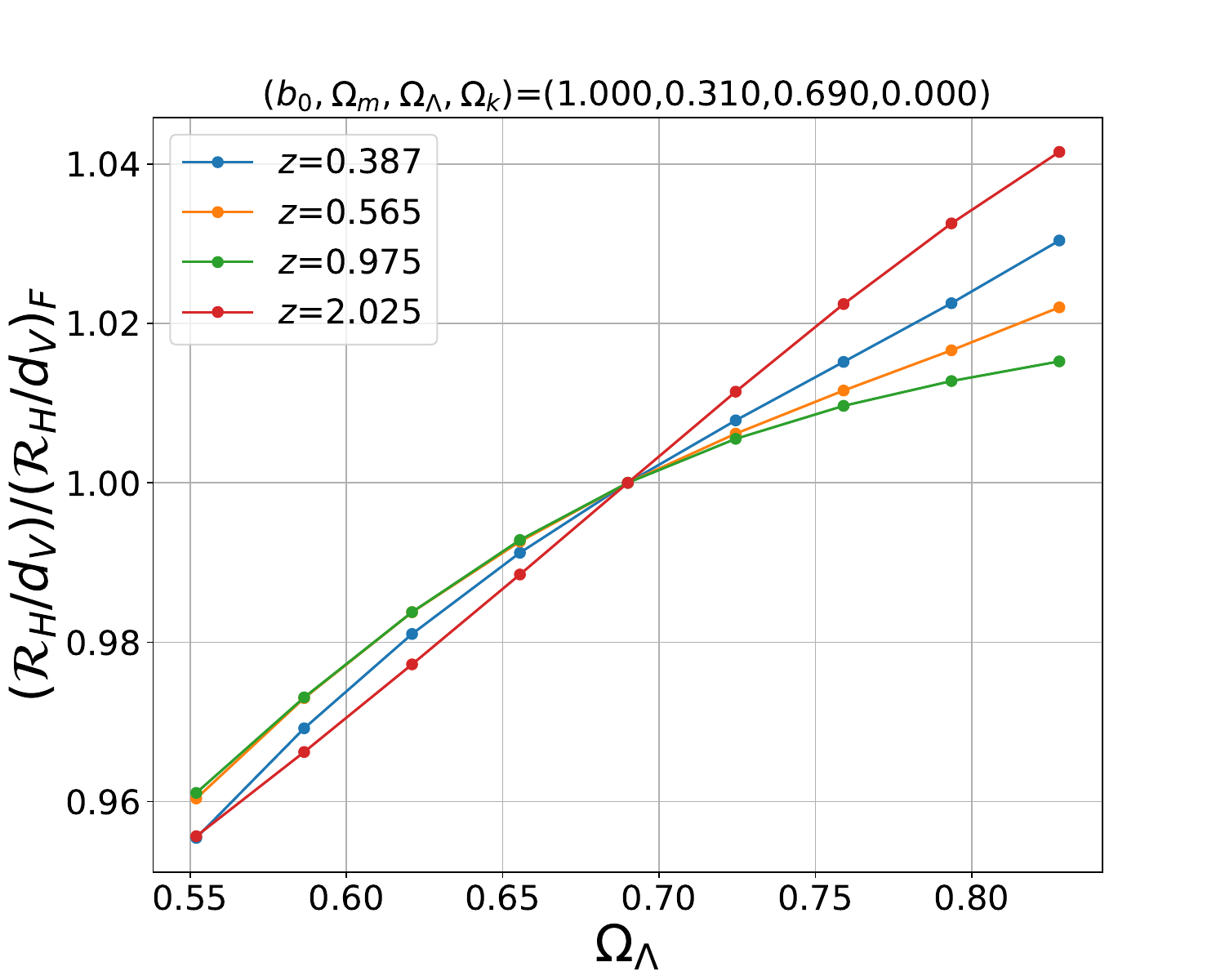}
	    	    \caption{\label{fig:determine_d2_b0_Om_OL_RhdV_div_Fiducial_Om_bfid10} Sensitivity of the normalised homogeneity scale, $\rh/d_V$, against the each values for a fiducial cosmology, $(b_0,\Omega_m,\Omega_{\Lambda})=(1.0,0.310,0.690)$. Different colours represent different redshift. [see \refS{sec:Sensitivity_of_the_normalised_homogeneity_scale}] }
	   \end{figure}%

In the left hand panel of \refF{fig:determine_d2_b0_Om_OL_RhdV_div_Fiducial_Om_bfid10}, we show the relationship between the normalised homogeneity scale and $\Omega_m$. We have applied the first bias model, $b_{1,\rh}(z)$, and the redshift space distortion (RSD) model.
The figure shows that around the fiducial cosmology the normalised homogeneity scale increases as a function of $\Omega_m$, with a variation of more than 10\%. The right hand panel of \refF{fig:determine_d2_b0_Om_OL_RhdV_div_Fiducial_Om_bfid10} shows that around the fiducial cosmology the normalised homogeneity scale increases as a function of $\Omega_{\Lambda}$ with a variation of more than 5\%. 
Different colour lines correspond to different redshift bins. 
These trends are true in all redshift bins. We also studied the case of a biased tracer and we reached the same conclusions.
Therefore, this shows that we can use the normalised homogeneity scale as a cosmological probe.

\subsection{MCMC set-up}\label{sec:MCMC}

We used an MCMC\footnote{We used the publicly available code, pymc \url{https://pymc-devs.github.io/pymc/}. } to sample the posterior probability distribution of our cosmological parameter space $p=(b_0,\Omega_m,\Omega_{\Lambda})$, to determine the cosmological constraints provided by the normalised homogeneity scale.
The likelihood for a given set of cosmological parameters, $p$, is expressed as $\mathcal{L}(p)\propto \exp\left[ - \chi^2(p)/2 \right]$, where $\chi^2$ is:
\begin{equation}\label{eq:chi2_final}
		\chi^2(p) = \Delta(p) \textbf{C}^{-1} \Delta^T (p) \; ,
\end{equation}
where $\Delta(p) = \textbf{O} - \textbf{M}(p)$; \textbf{M} is the theoretical prediction; \textbf{O} is our observable and $\textbf{C}$ is the covariance matrix which are given by \refEq{eq:data} and \refEq{eq:covmatrix}. 
 
In addition to our observable, we also used prior information on the $(\Omega_{m},\Omega_{\Lambda})$ plane as obtained by Planck 2018 \cite{aghanim2018planck} from the CMB and the CMB$+$Lensing measurements. We use the Ruby-Gelman \cite{gelman1992single} test as a convergence criterion for each MCMC realisation, $RG-1<0.01$.

	   \begin{figure}[h!]
	   \hspace{-1.cm} 
	    \includegraphics[width=180mm]{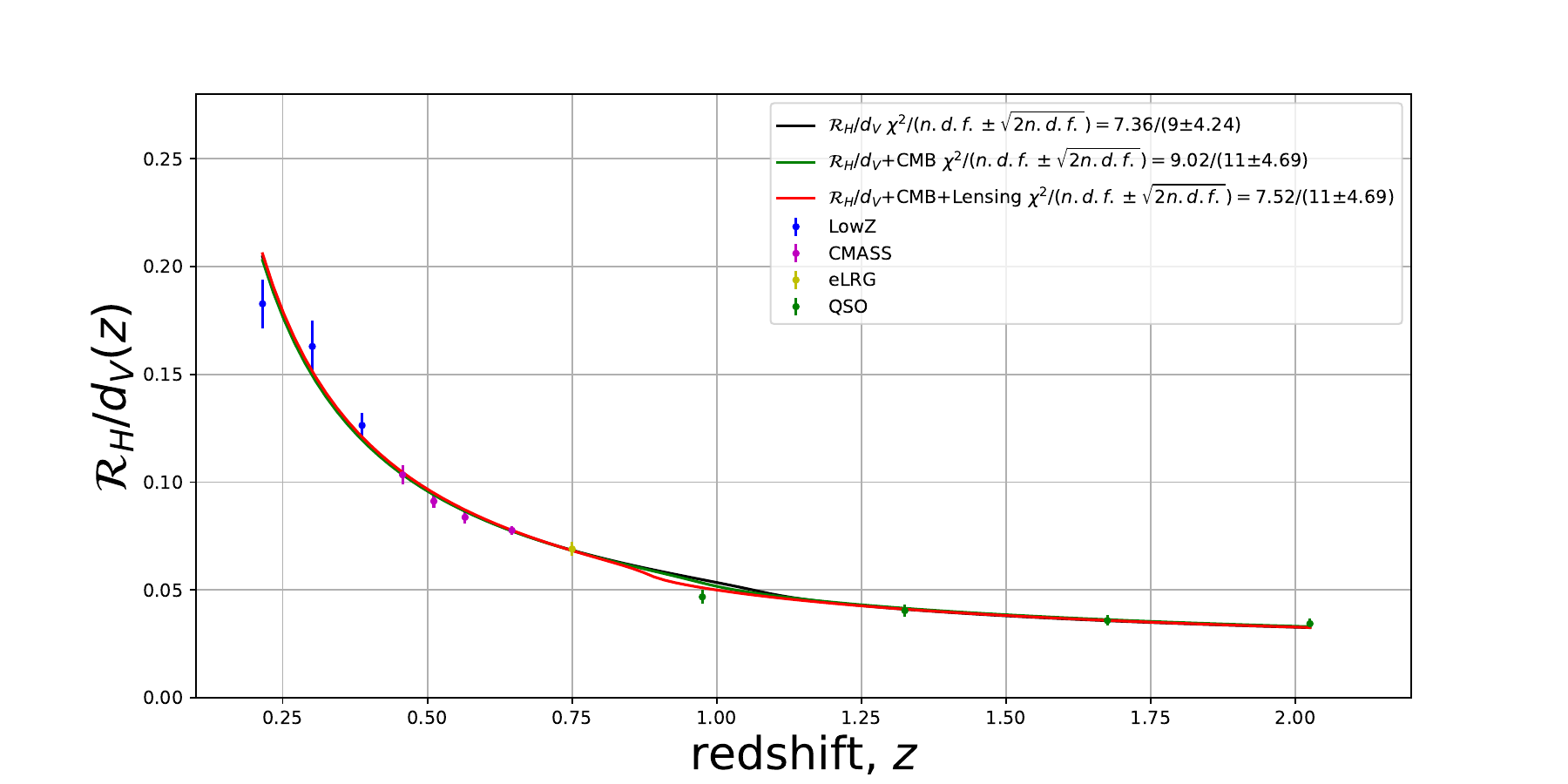}
	    
	    \caption{\label{fig:AMCMC_bias_model_Data_best_fit} Redshift evolution of the normalised homogeneity scale for the four data galaxy samples, LowZ(blue), CMASS (purple), eLRG (yellow) and QSO (green). The black line is the model that best fits the data alone. 
	    [see \refS{sec:Results}].}
	   \end{figure}%

\subsection{Results}\label{sec:Results}
In \refF{fig:AMCMC_bias_model_Data_best_fit}, we present the normalised homogeneity scale for the galaxy distribution of the universe as a function of redshift. The quantity, $\rh/d_V$, is plotted for the four galaxy samples that we study, i.e. the LowZ sample (blue), the CMASS sample (magenta), eLRG sample (yellow) and the QSO sample (green). Figure \ref{fig:AMCMC_bias_model_Data_best_fit} also shows the three best fitting models: one using only the galaxy data; the galaxy data combined with the CMB; and the galaxy data in combination with CMB$+$Lensing. 
Given our $\chi^2$, we find a very good agreement between the normalised homogeneity scale with the single bias parameter model and the data. Note that when we add the priors, we add 2 degrees of freedom. The normalised homogeneity scale increases with time since as the universe expands structures grow. 
At redshifts $z\simeq1.0$, there is a non-smooth variation of the model due to the BAO feature, known to be at a scale of $105 \hMpc$ \cite{eisenstein2005detection}, in the $\mathcal{D}_2(r)$ function.  This effect occurs at slightly different redshifts, for the different parameter values in $\rh/d_V$. We explore this effect in \refApp{sec:BAOfeatureOnD2}. 

\begin{figure}[ht!]
	   \hspace{-2.5cm} 
	    \includegraphics[width=200mm]{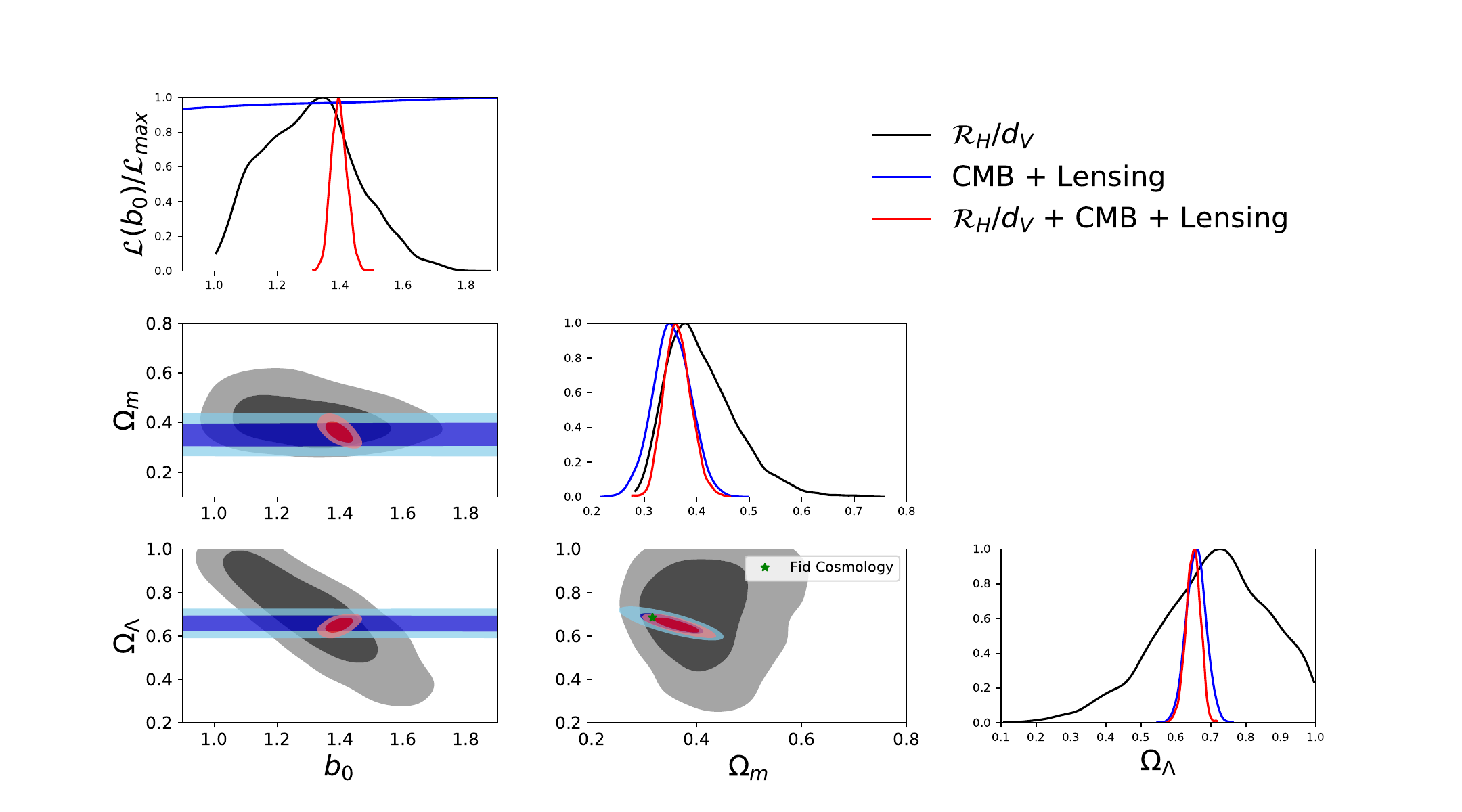}
	    
	    \caption{\label{fig:RH_CMBLensing_RHCMBLensing} Contours of 68\% (dark) and $95\%$ (light) C.L. of ($b_0,\Omega_m,\Omega_{\Lambda}$) planes using actual data, 
	    $\rh/d_V$ (black), 
	    CMB$+$Lensing (blue) and $\rh/d_V+$CMB$+$Lensing (red). The green star denotes the values of our fiducial cosmology. The diagonal panels show the normalised likelihood and the mean and the standard deviation of each parameter colour-coded for each probe. [See text for \refS{sec:Results}] }
	   \end{figure}

In \refF{fig:RH_CMBLensing_RHCMBLensing}, we present the result of our MCMC analysis for the open $\Lambda$CDM model. We show the marginalised contours of the 6 different combinations of the ($b_0,\Omega_m,\Omega_{\Lambda}$) planes for the $\rh/d_V$ alone (black), CMB$+$Lensing (blue), and the combination of $\rh/d_V+$CMB$+$Lensing (red). The green star denotes the values of the fiducial cosmology.   

The results for the probe combinations that we studied in this work, are shown in \refT{tab:Combination}. 
We find that $\rh/d_V$ alone can constrain the measurement of $\Omega_m$. The addition of information from $\rh/d_V$ improves constraints relative to the CMB alone with
40\% 
reduction on the uncertainty for $\Omega_m$ 
and 
34\% 
for $\Omega_{\Lambda}$. 
While when we add the normalised homogeneity scale to the CMB+Lensing we get an improvement of a
31\% 
reduction of the uncertainty for $\Omega_m$ 
and 
28\% 
for $\Omega_{\Lambda}$. 
These results show that the combination of the homogeneity scale with the CMB provides results comparable to CMB+Lensing.

Furthermore, we find that the bias values as obtained by the $\rh/d_V$ alone and using all galaxy samples at once, given in \refT{tab:Combination}, $b^{\rh/d_V}(z_{eff}=0.32) = 1.50\pm 0.22$ and $b^{\rh/d_V}(z_{eff}=0.57) = 1.62\pm 0.25$ are consistent at $2\sigma$ with the typical values of bias from the BOSS collaboration\cite{2017MNRAS.470.2617A} which are between $b(z_{eff}=0.32) \simeq b(z_{eff}=0.57) \simeq [1.95-2.1]$. Using the CMB and CMB+Lensing, the constraints of the bias values become more tight but they cannot be compared to the BOSS collaboration\cite{2017MNRAS.470.2617A} values. The BOSS bias values are obtained independently from using the CMB and CMB+Lensing constraints. When one uses a multivariate analysis to constrain the combination of bias with other values one constrains the bias parameter more. 

These results demonstrate that $\rh/d_V$ can be used as a probe to constrain cosmological parameters. In particular, it can be used to improve the cosmological measurements in the $(\Omega_m,\Omega_{\Lambda})$ plane\footnote{Our analysis is available under GNU licence \url{https://github.com/lontelis/CoHo2}. 
This is further evidence of the $\Lambda$CDM model, the standard model of cosmology. 
}. 

\begin{table}[h!]

\caption{\label{tab:Combination}  Mean and standard deviation of the measured cosmological parameters $(\Omega_m,\Omega_{\Lambda})$, using different combination of data. [See \refS{sec:Results}] }

		\hspace{-1.0cm} 
		\begin{tabular}{l|cccc} 
		\hline
		Probe Combinations & $b_0$ & $\Omega_m$ & $\Omega_{\Lambda}$ & $\chi^2\pm\sqrt{2 ndf},\ ndf=9$  \ \\
		\hline
		\hline
		$\mathrm{CMB}$ 	                            & -              & $0.473 \pm 0.089$ & $0.571 \pm 0.070$   & - \\
		$\mathrm{CMB}+\mathrm{Lensing}$    & -              & $0.352 \pm 0.036$ & $0.658 \pm 0.029$   & -    \\
		

$\mathcal{R}_H/d_V$ &
 $1.306 \pm 0.149$ &
 $0.413 \pm 0.068$ &
 $0.705 \pm 0.161$ &
 $7.362 \pm 4.243$
 \vspace*{0.0mm} \\
$\mathcal{R}_H/d_V$$+$CMB &
 $1.373 \pm 0.031$ &
 $0.411 \pm 0.053$ &
 $0.619 \pm 0.046$ &
 $9.015 \pm 4.690$
 \vspace*{0.0mm} \\
$\mathcal{R}_H/d_V$$+$CMB$+$Lensing &
 $1.397 \pm 0.026$ &
 $0.363 \pm 0.025$ &
 $0.650 \pm 0.021$ &
 $7.524 \pm 4.690$
 \vspace*{0.0mm} \\
 \hline 
 \end{tabular}

\end{table}

\subsection{Study of systematic effects}\label{sec:NullTests}

In order to quantify any bias coming from the values of the fiducial cosmology, we performed two dedicated studies. Firstly, we repeated the measurement on the data using a different fiducial cosmology. The two different cosmologies are defined by \refEq{fid-cosmo} (denoted by $p_{F}$) and \refEq{eq:fid-cosmology-test} (denoted by $p_{F2}$). These cosmologies differ from each other by $15\%$ for $\Omega_m$ and by $6\%$ for $\Omega_{\Lambda}$.  We present the results in \refT{tab:Systematic_fiducial}. The systematic bias obtained by the different fiducial cosmologies is not significant with respect to the statistical error. Ideally, this study should be performed on mock catalogues, but mocks are not publicly available in all redshift bins.

Secondly, in \refApp{sec:Validation_test_data_mock_LowZCMASSeLRG}, we present further test in the subset of the redhift bins, $z<0.8$, where the mock catalogues are available. Using the mock catalogues and even though this test-measurement is less precise than the measurement at the whole redshift bin $0.172 \leq z \leq 2.2$, we retrieve the fiducial cosmology within $98\%$ percentile, which validates our modelling and our analysis.

In conclusion since the constraints from the data already agree at less than $1\sigma$ level from using two different fiducial cosmologies that differ from each other more than $1\sigma$ level our analysis is validated.

\begin{table}[h!]

 \caption{\label{tab:Systematic_fiducial}  Systematic study on the measurement. Combination of probes that measure the model parameters using bias model 1, \refEq{eq:bias_model2} and two different fiducial cosmologies. [See \refS{sec:NullTests}] 
  }

\hspace{-1cm}
\begin{tabular}{l | c c c c} \hline
$b_{1,\rh}(z)$ & $b_0$ & $\Omega_m$ & $\Omega_{\Lambda}$ & $\chi^2\pm\sqrt{2.ndf}$, $ndf=9$ \\ 
 \hline
  \hline
 \hline
$p_{F}$-$\mathcal{R}_H/d_V$ &
 $1.306 \pm 0.149$ &
 $0.413 \pm 0.068$ &
 $0.705 \pm 0.161$ &
 $7.362 \pm 4.243$
 \vspace*{0.0mm} \\
$p_{F,2}$-$\mathcal{R}_H/d_V$ &
 $1.348 \pm 0.157$ &
 $0.427 \pm 0.078$ &
 $0.643 \pm 0.183$ &
 $7.456 \pm 4.243$
 \vspace*{0.0mm} \\
$p_{F}$-$\mathcal{R}_H/d_V$$+$CMB &
 $1.373 \pm 0.031$ &
 $0.411 \pm 0.053$ &
 $0.619 \pm 0.046$ &
 $9.015 \pm 4.690$
 \vspace*{0.0mm} \\
$p_{F2}$-$\mathcal{R}_H/d_V$$+$CMB &
 $1.371 \pm 0.033$ &
 $0.411 \pm 0.057$ &
 $0.618 \pm 0.047$ &
 $8.891 \pm 4.690$
 \vspace*{0.0mm} \\
$p_{F}$-$\mathcal{R}_H/d_V$$+$CMB$+$Lensing &
 $1.397 \pm 0.026$ &
 $0.363 \pm 0.025$ &
 $0.650 \pm 0.021$ &
 $7.524 \pm 4.690$
 \vspace*{0.0mm} \\
$p_{F2}$-$\mathcal{R}_H/d_V$$+$CMB$+$Lensing &
 $1.393 \pm 0.024$ &
 $0.364 \pm 0.025$ &
 $0.649 \pm 0.020$ &
 $7.482 \pm 4.690$
 \vspace*{0.0mm} \\
 \hline 
 \end{tabular}

    \end{table}


\section{Conclusion and discussion}\label{sec:conclusion}

In this work, we have demonstrated that the normalised characteristic scale of transition to cosmic homogeneity, $\rh/d_V$, can be used as a cosmological probe with large scale structure surveys. For this, we have used four publicly available galaxy samples, LowZ, CMASS, eLRG and QSO of the Sloan Digital Sky Survey. We have also used an empirical approach in order to extract a redshift dependent bias model for the normalised homogeneity scale at all redshifts from the different galaxy samples. 

In order to quantify the additional cosmological information contained in the normalised homogeneity scale, 
we have performed an MCMC analysis and we have explored the open \LCDMn\ model. 
By combining our measurements with prior information from CMB+Lensing, we found 
 \-$\Omega_{m} =0.363 \pm 0.025$
and
\-$\Omega_{\Lambda}= 0.650 \pm 0.021$,
consistent with a flat \LCDMn\ cosmological model at the $1\sigma$ level. The inclusion of $\rh/d_V$ improves CMB$+$Lensing constraints alone by a reduction on the uncertainty of 
31\% 
for $\Omega_m$ 
and 
28\% 
for $\Omega_{\Lambda}$. There is, therefore, a clear gain when it is combined with CMB+Lensing information.

The normalised homogeneity scale shows evidence for a flat-$\Lambda$CDM cosmology. This is in agreement with current literature on the combination of galaxy clustering and other probes \cite{aghanim2018planck}. In particular, we find 
$\Omega_k^{\mathrm{CMB}+\mathrm{Lensing}+\mathrm{\rh/d_V}}=-0.0126\pm0.0078$
which is comparable at $2\sigma$ with the Planck value,
$\Omega_k^{\mathrm{CMB}+\mathrm{Lensing}+\mathrm{BAO}}=0.0007\pm0.0019$. The BAO analysis performed in \citet{aghanim2018planck}, takes into account two dimensional information from galaxy clustering, $r_{\perp},r_{||}$. In contrast, in this work, we have not taken that into account, which might result in our lower constraining power.
Therefore, further studies are required with more sophisticated analysis to combine this measurement with other probes.

In this work, we measured the homogeneity scale on the QSO sample independently from \citet{Laurent} and \citet{2018arXiv180911125G}. We acquired results that are compatible and more precise. We have more precise results than \citet{2018arXiv180911125G}, since they used narrower redshift bins than us.  \citet{Laurent} have used an outdated QSO catalogue from BOSS, while we are using the eBOSS QSO catalogue which is both deeper and denser. Therefore, we get more precise measurements.


\citet{2019arXiv190102400N} have argued that the homogeneity scale cannot be considered as a standard ruler and that it cannot constrain cosmological parameters since it does not have a one-to-one dependence on $\Omega_m$. We agree with the first statement. Since $\rh$ evolves with time (and therefore redshift) it cannot be considered to be a standard ruler. However, we disagree with their second conclusion. In this paper, we have shown that the normalised homogeneity scale, $\rh/d_V$, without the addition of other probes, can be used  to place a constraint on $\Omega_m$. This is due to the fact that even though the dependence of the homogeneity scale on $\Omega_m$ is weak, the dependence of the homogeneity scale normalised with the volume distance is much stronger as we have shown in \refS{sec:Sensitivity_of_the_normalised_homogeneity_scale} . In addition, we have shown that in combination with CMB$+$Lensing, the normalised homogeneity scale also improves the constraint on $\Omega_{m}$ and $\Omega_{\Lambda}$.

In conclusion, we have revealed the complementarity of the homogeneity scale with respect to other cosmological probes. 

Finally, we stress that this analysis can be performed and improved upon in the light of more observational data from current and future survey such as SDSS-IV\cite{dawson2016sdss}, DESI\cite{aghamousa2016desi}, Euclid\cite{2016arXiv160600180A} and LSST\cite{2009arXiv0912.0201L}. Furthermore, analogous methods could be applied to data from SKA\cite{dewdney2009square}. A similar analysis can also be applied by measuring the normalised homogeneity in the temperature fluctuations of CMB as observed by Planck \cite{aghanim2018planck}. Potentially, one could investigate additional observational systematic effects on our probe \cite{ntelisFuture}, but as shown in \cite{PIERROS_THESIS} the known systematics (modelled by the weights), do not affect the measurement of the homogeneity scale at the $1\sigma$ level. One can also extend this analysis to test Modified Gravity models or Effective Field Theory models\cite{2016arXiv160600180A}. We leave these analyses for future work. 


\vspace{0.5cm}
\hspace{-1cm} \textbf{AKNOWLEDGEMENTS}\\
	
	We would like to thank Christian Marinoni, Julien Bel, Jean-Marc Le Goff, James Rich, Jean-Christophe Hamilton, Adam Morris and Francis Bernardeau for useful suggestions and discussions. We would like to thank the two anonymous referees for their fruitful comments that help improve this study. We acknowledge open libraries support \cite{Hunter:2007,Walt:2011:NAS:1957373.1957466,2019arXiv190710121V}.
	
	
	PN is funded by  'Centre National d'\'etude spatiale' (CNES), for the Euclid project.
	
	AJH acknowledges the financial support of the OCEVU LABEX (Grant No. ANR-11- LABX-0060) and the A*MIDEX project (Grant No. ANR-11-IDEX- 0001- 02) funded by the Investissements dAvenir french government program managed by the ANR as well as the Centre National de la Recherche Scientifique (CNRS).

	 This work also acknowledges support from the ANR eBOSS project (under reference ANR-16-CE31-0021) of the French National Research
Agency.	
	
	
	This research used resources of the IN2P3/CNRS and the Dark Energy computing Center funded by the OCEVU Labex (ANR-11-LABX-0060).
	
	The French Participation Group of SDSS was supported by the French National Research Agency under contracts ANR-08-BLAN-0222, ANR-12-BS05-0015-01 and ANR-16-CE31-0021.
	
	Funding for SDSS-III has been provided by the Alfred P. Sloan Foundation, the Participating Institutions, the National Science Foundation, and the U.S. Department of Energy Office of Science.
The SDSS-III web site is http://www.sdss3.org/.
	
	SDSS-III is managed by the Astrophysical Research Consortium for the Participating Institutions of the SDSS-III Collaboration including the University of Arizona, the Brazilian Participation Group, Brookhaven National Laboratory, Carnegie Mellon University, University of Florida, the French Participation Group, the German Participation Group, Harvard University, the Instituto de Astrofisica de Canarias, the Michigan State/Notre Dame/JINA Participation Group, Johns Hopkins University, Lawrence Berkeley National Laboratory, Max Planck Institute for Astrophysics, Max Planck Institute for Extraterrestrial Physics, New Mexico State University, New York University, Ohio State University, Pennsylvania State University, University of Portsmouth, Princeton University, the Spanish Participation Group, University of Tokyo, University of Utah, Vanderbilt University, University of Virginia, University of Washington, and Yale University.

\appendix
\section*{Appendices}
\addcontentsline{toc}{section}{Appendices}
\renewcommand{\thesubsection}{\Alph{subsection}}

\subsection{Fiducial quantities}\label{sec:Fiducial_cosmologies}
In \refT{tab:Reference}, we summarise the fiducial cosmology values and prior measured values used in this study.

\begin{table}[h!]

   \caption{\label{tab:Reference}  Reference table for the ($\Omega_m,\Omega_{\Lambda}$)-plane. The first two lines show the fiducial cosmology used in this work, the last two lines show the cosmology measurements as obtained by Planck2018\cite{aghanim2018planck}. }

\hspace{1.2cm}
\begin{tabular}{l | c c c c} \hline
 & $b_0$ & $\Omega_m$ & $\Omega_{\Lambda}$  \\ 
 \hline
Data - Fiducial ($p_F$) &
 $- $ &
 $0.316 $ &
 $0.684$

\vspace*{0.0mm} \\
 
Data - Fiducial 2 ($p_{F2}$) &
 $- $ &
 $0.274$ &
 $0.726$
\vspace*{0.0mm} \\
 
 Mock - QPM ($p_{qpm}$) &
 $- $ &
 $0.290$ &
 $0.710$
 \vspace*{0.0mm} \\
 
CMB (Planck 2018) &
 $- $ &
 $0.473 \pm 0.089$ &
 $0.571 \pm 0.072$
 \vspace*{0.0mm} \\
CMB+Lensing(Planck 2018) &
 $ - $ & 
 $0.352 \pm 0.036$ &
 $0.658 \pm 0.029$
 \vspace*{0.0mm} \\
 \hline 
 \end{tabular}

\end{table}

\subsection{Bootstrap covariance validation}\label{sec:Bootstrap_covariance_validation}
\begin{figure}[h!]
	   \hspace{-0.0cm} 
	   \vspace{-0.3cm}
	    \includegraphics[width=0.325\linewidth, keepaspectratio]{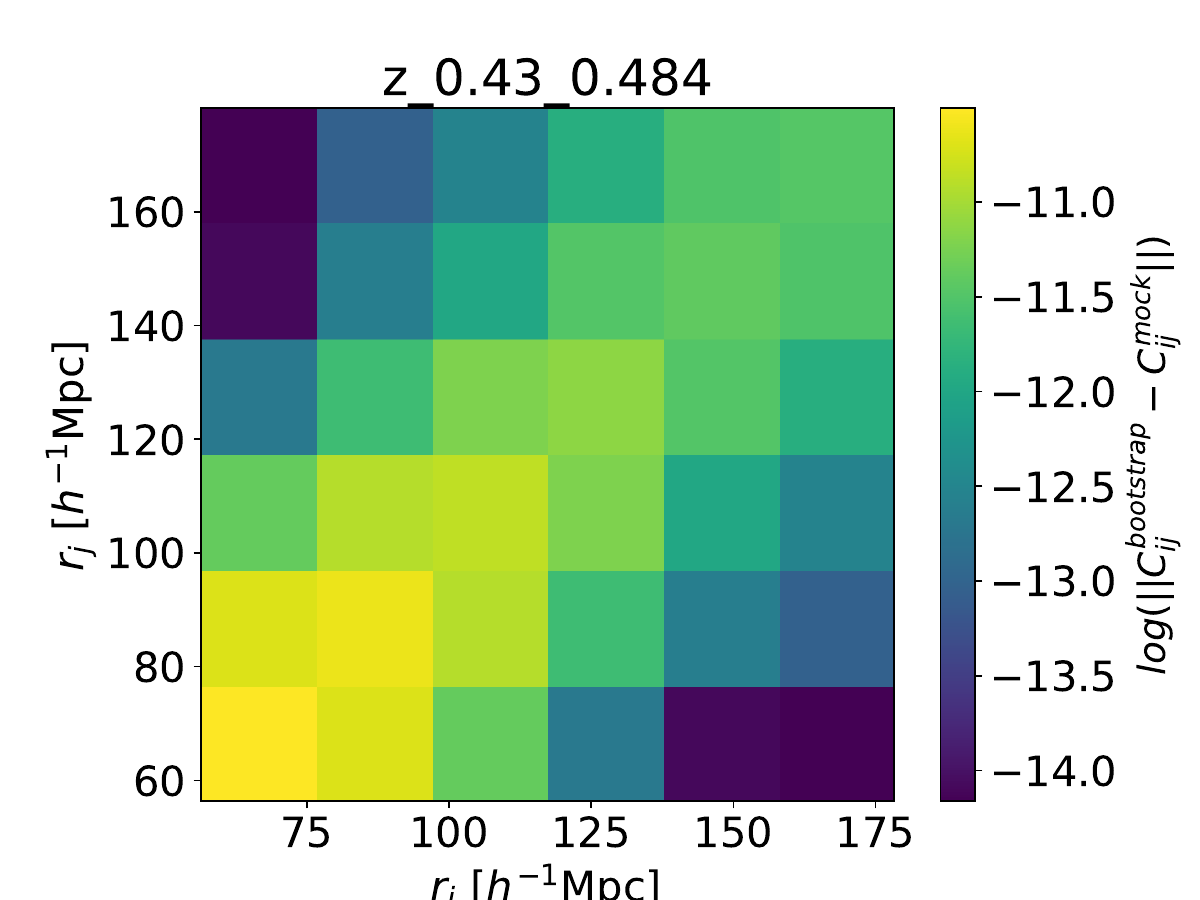}
	    \includegraphics[width=0.325\linewidth, keepaspectratio]{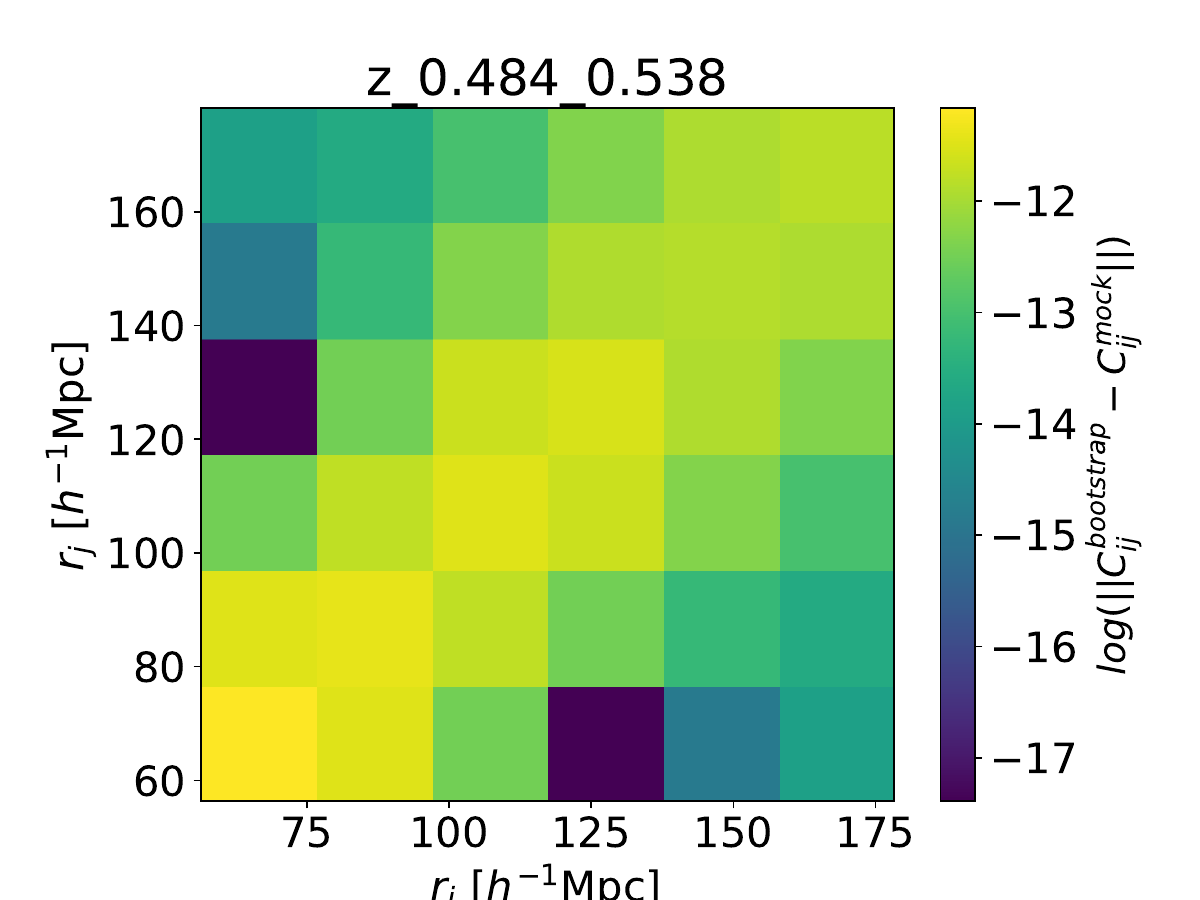}
	    \includegraphics[width=0.325\linewidth, keepaspectratio]{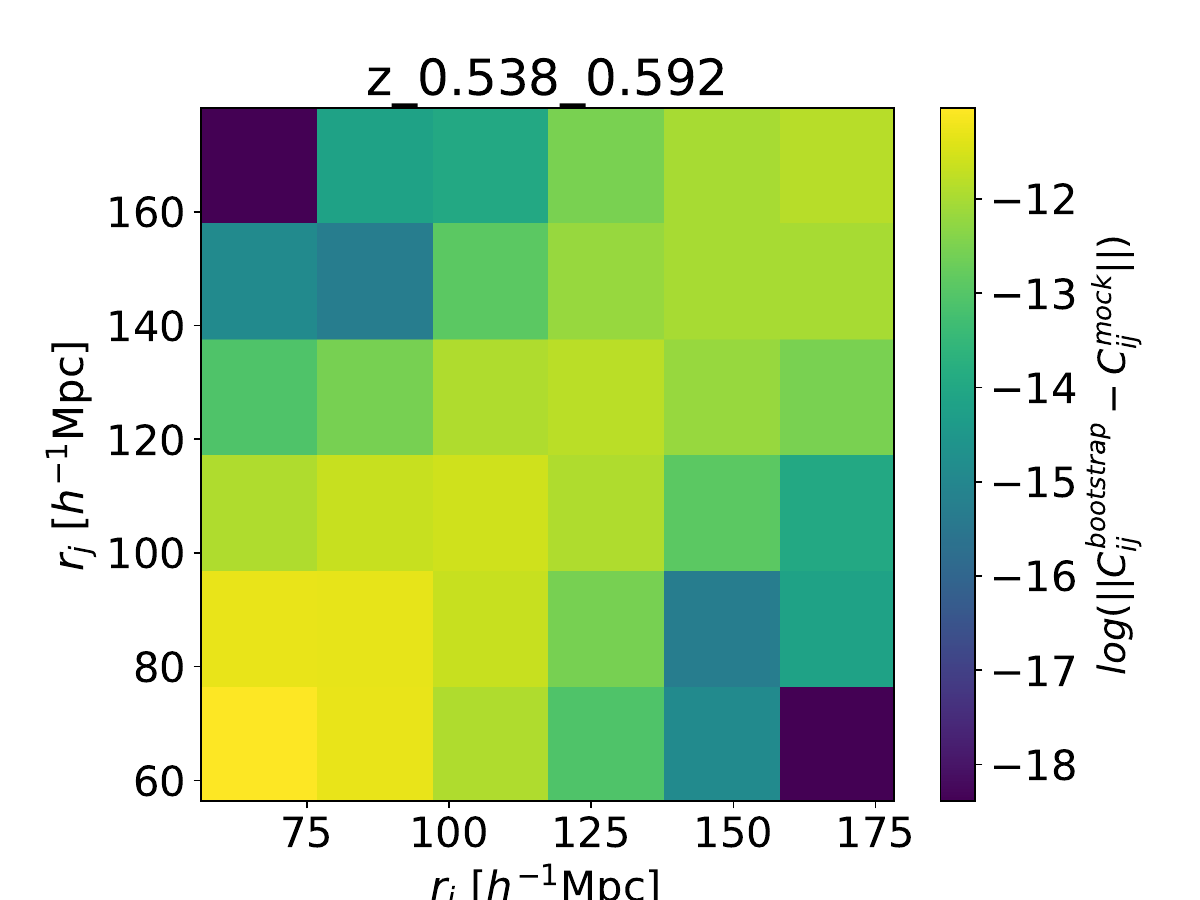}
	    	    \caption{\label{fig:Calculate_Cov_ratio_mock_bootstrap_CMASS_z} Absolute difference of the covariance matrices in logarithmic scale for the fractal dimension, $\Dd{}(r)$, between the bootstrap method and the mock method for the 3 redshift bins of the CMASS galaxy catalogues. [see \refApp{sec:Bootstrap_covariance_validation}] }
	   \end{figure}

We study the Covariance matrix for the fractal dimension, $\Dd{}(r)$, as given by \refEq{eq:Covariance_Fractal}. We calculate the covariance for the CMASS galaxy sample for the first 3 redshift bins of the CMASS galaxy catalogue using the 100 mock catalogues, described in \refS{sec:mocks}. Then, we calculate the covariance matrix for the same galaxy catalogues using the bootstrap method. In \refF{fig:Calculate_Cov_ratio_mock_bootstrap_CMASS_z}, we give the absolute difference between the covariance matrix using the bootstrap method and the covariance matrix from mock catalogues. 

This result shows that the use of the bootstrap method results in a covariance matrix that approximates the constructed ones using the mock catalogues. This validates the use of the bootstrap method on the samples where we lack mock catalogues. In future work, we plan to use mock catalogues in order to update and improve this study further.

\subsection{Correlation matrix $\rho_{\rh/d_V}(z_i,z_j)$}\label{sec:Correlation_matrix_rhdv}
\begin{figure}[h!]
	   \hspace{-0.0cm} 
	   \vspace{1.0cm}
	    \includegraphics[width=0.325\linewidth, keepaspectratio]{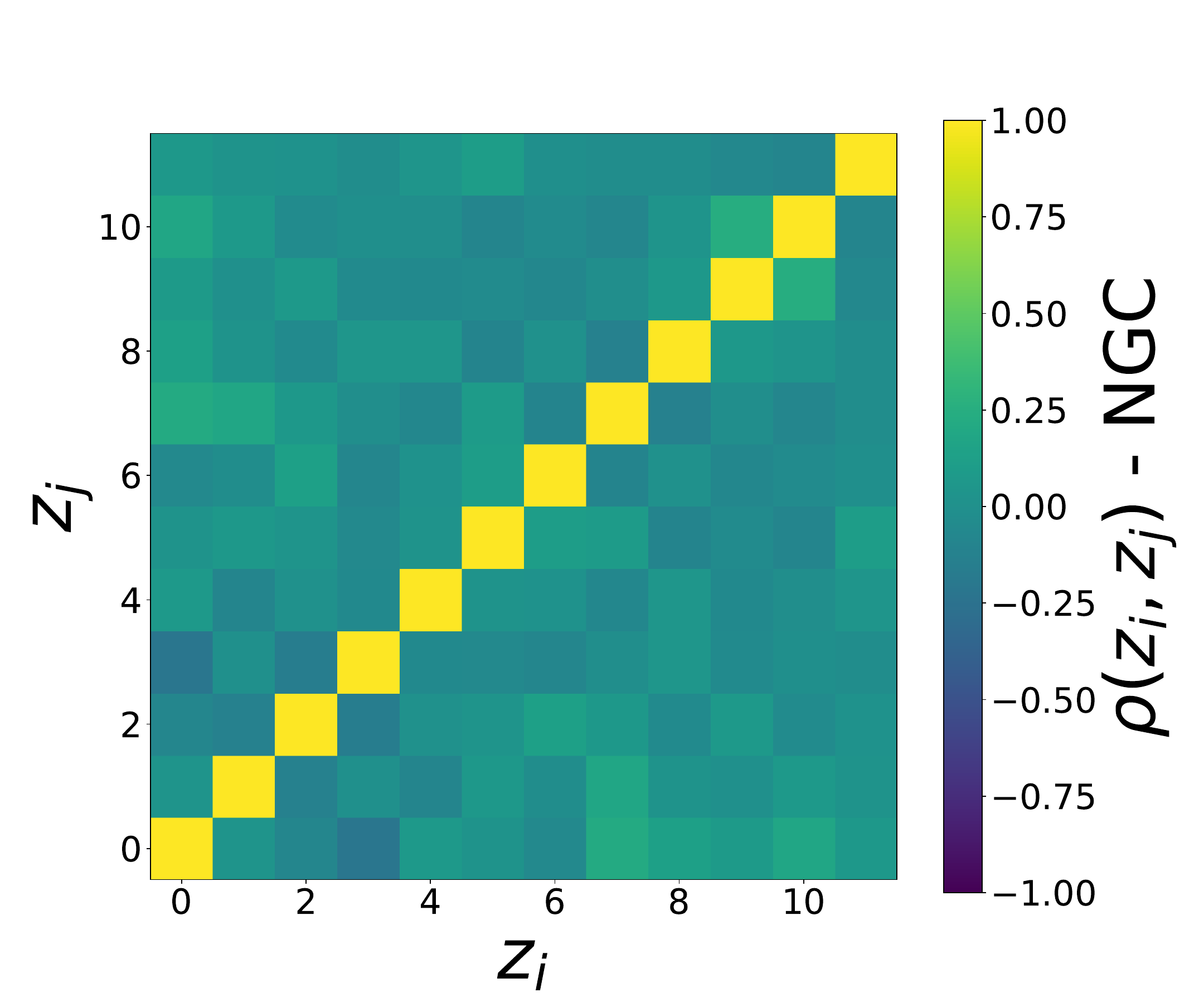}
	    \includegraphics[width=0.325\linewidth, keepaspectratio]{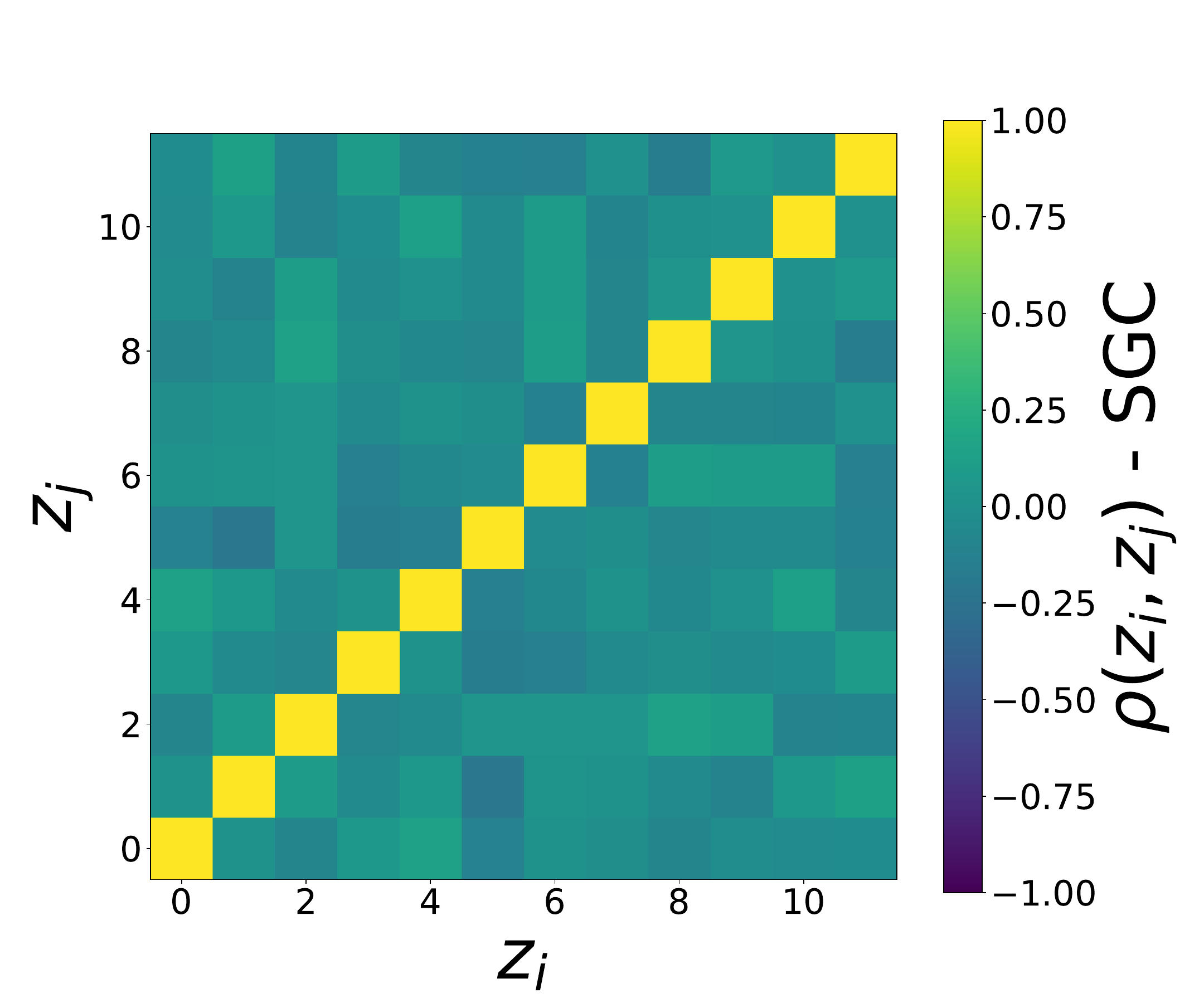}
	    \includegraphics[width=0.325\linewidth, keepaspectratio]{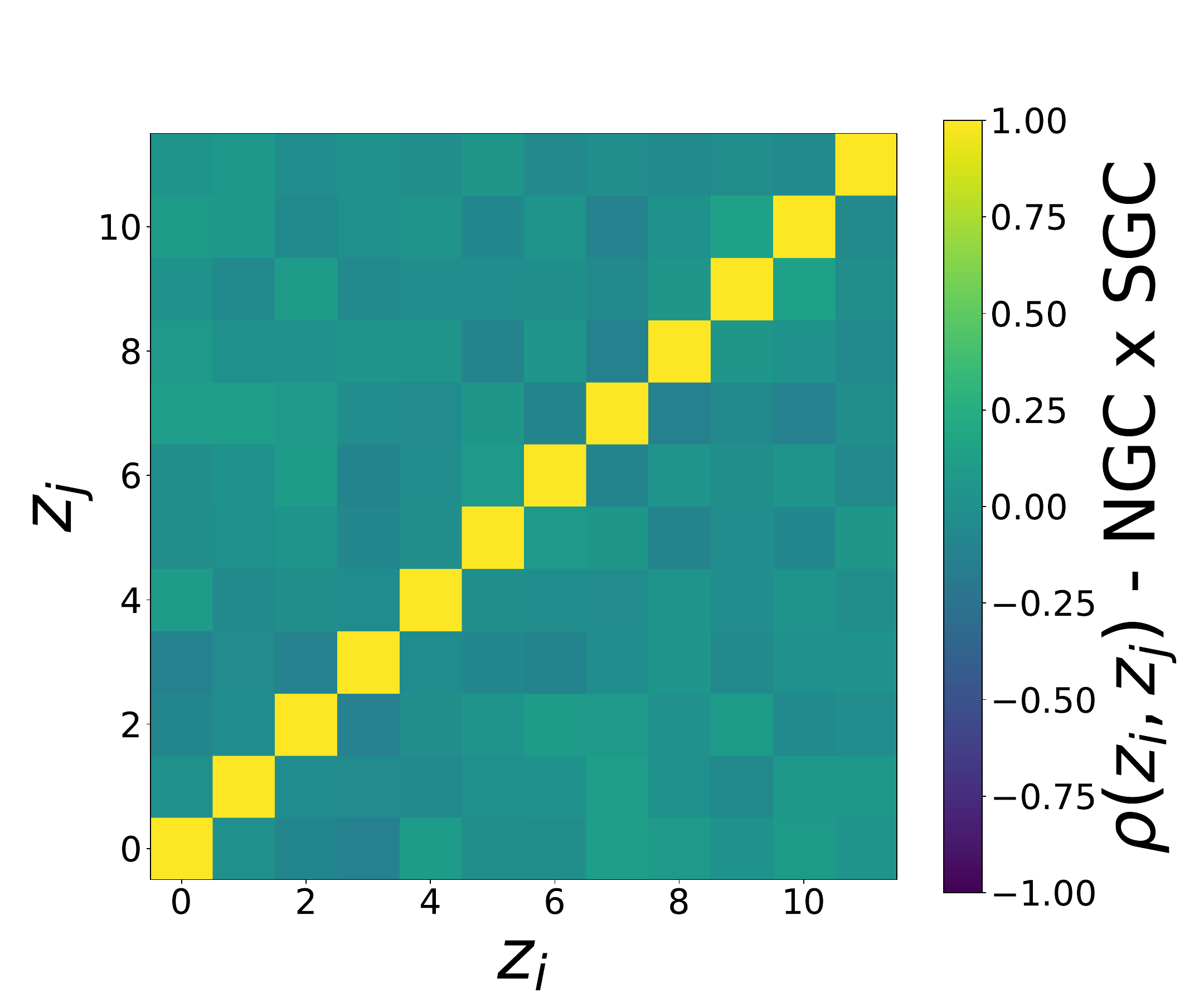}
	    \vspace{-1.5cm}
	    	    \caption{\label{fig:Calculate_Corrmatrix_of_RhdV} Correlation matrices of the normalised homogeneity scale as a function of redshift, $\rho_{\rh/d_V}(z_i,z_j)$ for the North(left), South(center) and Combination(right) of Galactic Caps. [see \refApp{sec:Observable_Estimation}] }
	   \end{figure}

We study the Correlation matrix of $\rh/d_V$ as a function of redshift for the North, South and the combination of the Galactic Caps. We use the combination of the galactic caps in our estimation of the weighted $\rh/d_V$ and the combination of the covariance as explained in \refS{sec:Observable_Estimation}. The covariance matrices for the North and South Galactic Caps are used to estimated the weighted average of $\rh/d_V$, while the combined covariance matrix is used for the fits with the theoretical model.  In \refF{fig:Calculate_Corrmatrix_of_RhdV}, we give the ratio of the matrix using the bootstrap method and the one using the mock catalogues for the three redshift bins. This result shows the necessity of taking into account the correlation of our observable from the different redshift bins. 

	   \begin{figure}[t!]
	    \includegraphics[width=0.499\linewidth, keepaspectratio]{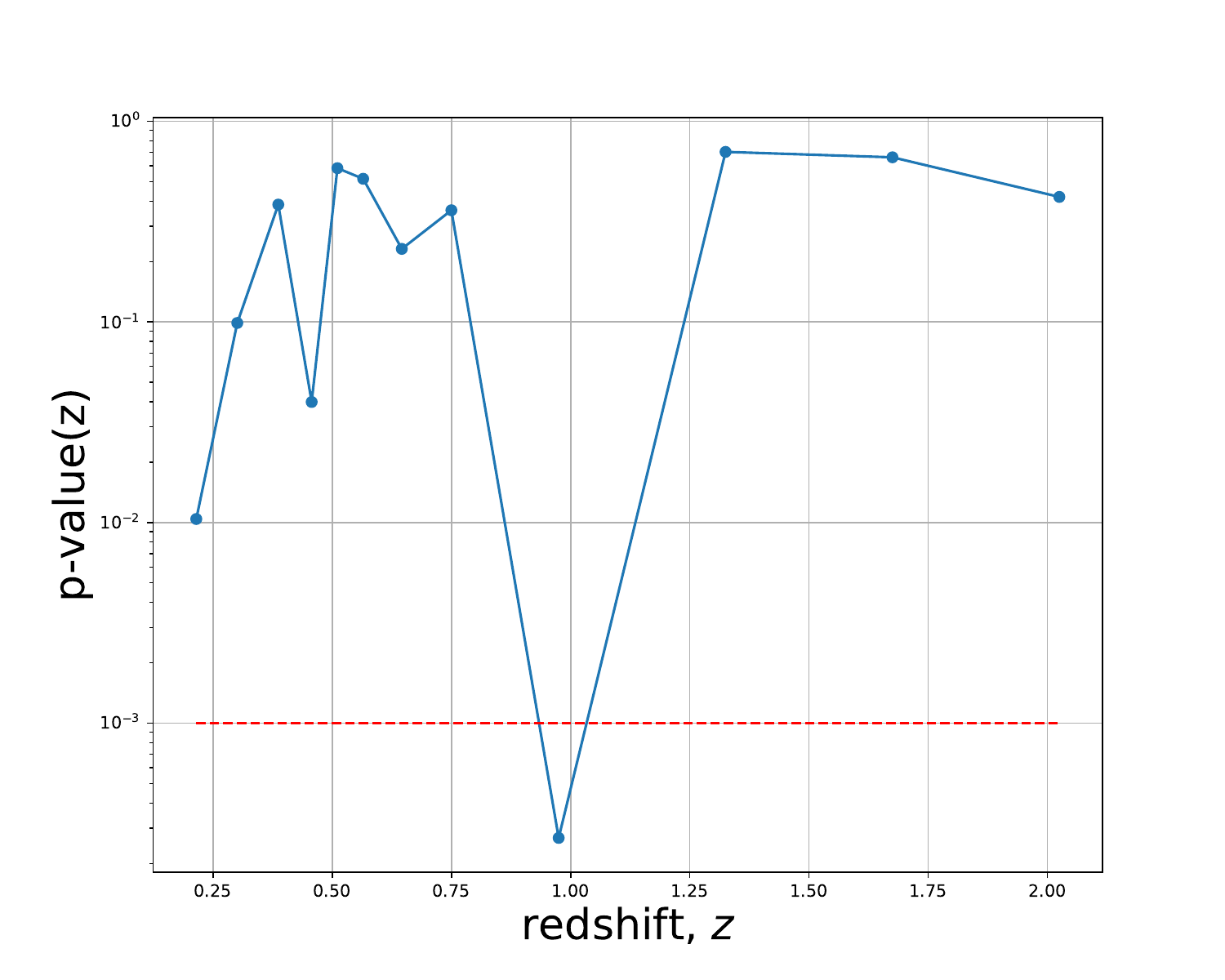}
	    \includegraphics[width=0.499\linewidth, keepaspectratio]{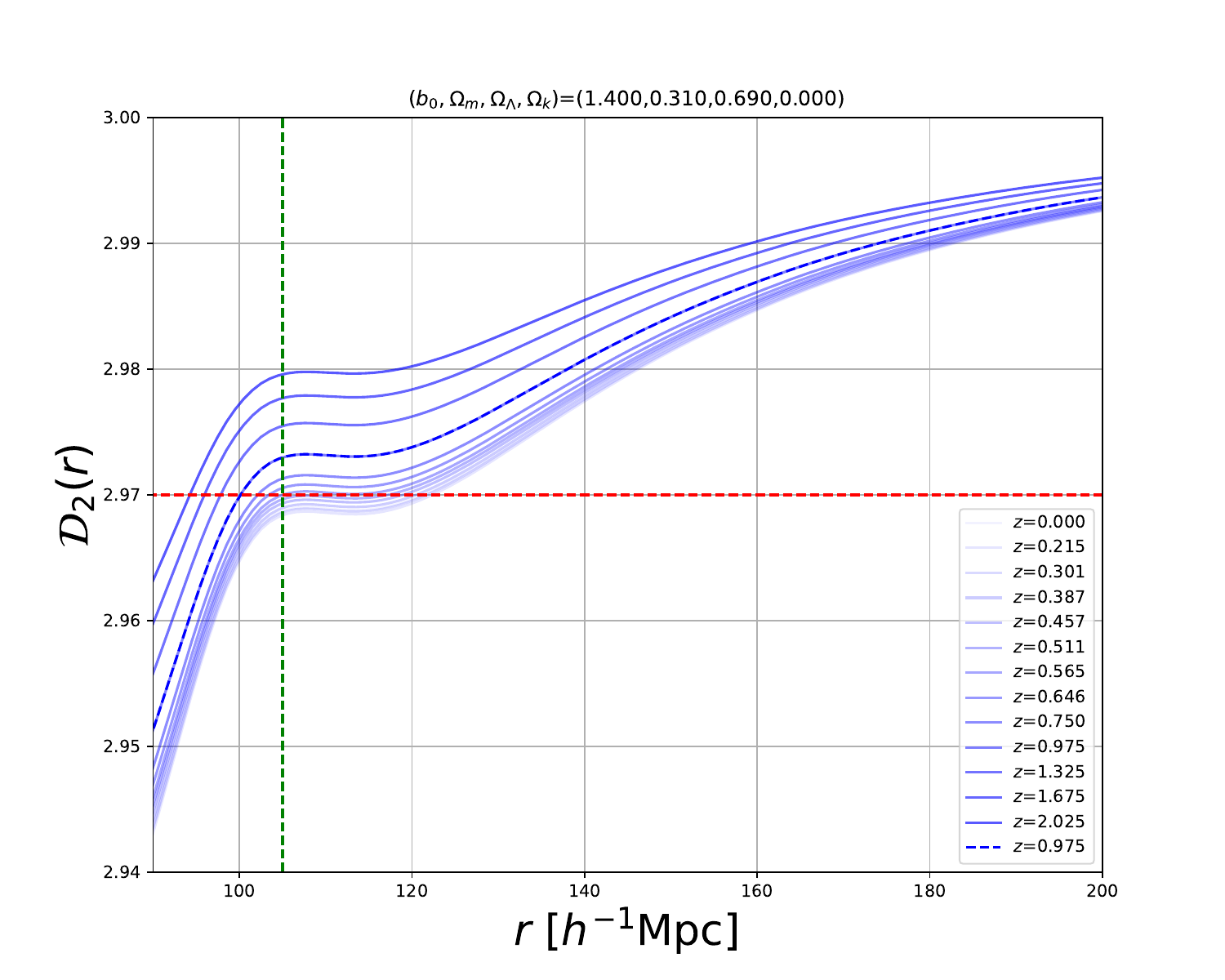}
	    \vspace{-1.0cm} 
	    	    \caption{\label{fig:determine_d2_r_z} 
		    \textit{Left:}
		     Omnibus normality test for the measurements of the normalized homogeneity scale, $\rh/d_V$ at each redshift bin, represented with the p-value at each redshift. 
		    [see \refApp{sec:likelihood_of_measurement}]
		    \textit{Right:} BAO feature on the Fractal dimension as a function of scales, $\mathcal{D}_2(r)$. Increasing shade represent increasing redshift.  The horizontal red dashed line represent the threshold for $1\%$ of homogeneity. The vertical green dashed line represents the BAO scale. [see \refApp{sec:BAOfeatureOnD2}] }
		    
	   \end{figure}%

\subsection{Normality test of the  $\rh/d_V$-measurement}\label{sec:likelihood_of_measurement}

In this section, we present an omnibus normality test\cite{10.1093/biomet/58.2.341} of the likelihood of the measurement of the normalised homogeneity scale, $\rh/d_V(z)$, at each redshift using the mocks. 
We show the results on the left panel in 
\refF{fig:determine_d2_r_z}
, where the red dash line shows the threshold above which the measurement at each redshift should be assumed to be Gaussian. We find that only one of the redshift bins does not follow a Gaussian distribution and this is reasonable since this redshift bin has the lowest number of galaxies. However, future surveys are going to reveal more galaxies in these redshift regions and we will be able to have a better study in the future. These results show that even though we have somewhat small redshift bins the Gaussian approximation is reasonable for almost every redshift bin.

\subsection{Effect of the BAO feature on the fractal dimension}\label{sec:BAOfeatureOnD2}
We demonstrate that the BAO feature is apparent in the fractal dimension at scales $105 \hMpc$. On the right panel, in \refF{fig:determine_d2_r_z}, we present the fractal dimension as a function of scale, $\mathcal{D}_2(r)$. Increasing shade represents increasing redshift. The horizontal red dashed line represent the threshold for $1\%$ homogeneity. The vertical green dashed line represents the BAO scale. The blue dashed line represents the fractal dimension as a function of scales for redshift $z=0.975$. It is obvious that the BAO feature results in a non-smooth behaviour of redshift dependence of the normalised homogeneity scale at redshifts $z\simeq 1.0$. To completely avoid the BAO feature one could define the threshold for the homogeneity scale to be less than $1\%$, for example $0.1\%$. However, \citet{ntelis2017exploring} have shown that a precision of this kind will be decreasing due to the fact that the  slope of the Fractal Dimension decreases. One could also choose a threshold larger than $1\%$, for example $2\%$. However, at theses scales the Redshift Space Distortion model becomes more complicated and additionally it would require some non-linear modelling of the Power Spectrum. Therefore, we leave this study for future work.

\subsection{Model validation test for data $z<0.8$}\label{sec:Validation_test_data_mock_LowZCMASSeLRG}

To validate our modelling, we use data and mocks in the redshift region $z<0.8$ where QPM mock catalogues are accessible to us. In particular, we perform the analysis as described in \refS{sec:NullTests} but now for three different cosmologies, $(p_F, p_{F,2}, p_{qpm})$, for both data and mock catalogues. In this case study, we take the mean and $1\sigma$ of the mocks because we want to simulate the error in the data which have as an error the $1\sigma$ of the mocks. In figure \refT{tab:Validation_test_data_mock_LowZCMASSeLRG} we show the results for both data and mock catalogues. In \refT{tab:Validation_test_data_mock_LowZCMASSeLRG_mocks_only} but now divided by the fiducial cosmology, i.e. $(\Omega_{m,qpm}, \Omega_{\Lambda,qpm})$. The information is degrade first of all because we provide the $98\%$ percentile and not the $68\%$ percentile of ($1\sigma$) quoted by our measurement. Furthermore, we consider only half of our data points, which is 7 data points instead of 12.
Finally, even though this measurement is less precise than the measurement at the whole redshift bin $0.172 \leq z \leq 2.2$, we retrieve the fiducial cosmology within $98\%$ percentile, which validates our modelling and our analysis. 

\begin{table}[h!]
\centering
   \caption{\label{tab:Validation_test_data_mock_LowZCMASSeLRG}  Validation test for data below $z<0.8$. The first column shows the configuration of the measurement of cosmogical parameters ("D" for data "M" for mock catalogues) with different fiducial cosmologies ($p_F$, $p_{F2}$, $p_{qpm}$ correspond to equations \ref{fid-cosmo}, \ref{eq:fid-cosmology-test}, \ref{qpm-cosmo}, respectively). Each of the 2-4 columns shows the mode of the variable of each chain and each $98\%$ percentile. The last column shows the $\chi^2$ test in respect of the $n.d.f$ }

\hspace{1.2cm}
\begin{tabular}{l c c c c} \hline
$\mathcal{R}_H/d_V$  & $b_0$ & $\Omega_m$ & $\Omega_{\Lambda}$ & $\chi^2\pm\sqrt{2.ndf}$, ndf=5 \\ 
 \hline
D-$p_F$ &
 $1.526 \pm 0.440$ &
 $0.609 \pm 0.466$ &
 $0.980 \pm 0.630$ &
 $11.377 \pm 3.162$
 \vspace*{0.0mm} \\
D-$p_{F2}$ &
 $1.543 \pm 0.483$ &
 $0.389 \pm 0.413$ &
 $1.080 \pm 0.638$ &
 $10.135 \pm 3.162$
 \vspace*{0.0mm} \\
D-$p_{qpm}$ &
 $1.557 \pm 0.446$ &
 $0.510 \pm 0.410$ &
 $0.990 \pm 0.630$ &
 $11.475 \pm 3.162$
 \vspace*{0.0mm} \\
M-$p_F$ &
 $1.827 \pm 0.484$ &
 $0.462 \pm 0.328$ &
 $1.080 \pm 0.556$ &
 $3.731 \pm 3.162$
 \vspace*{0.0mm} \\
M-$p_{F2}$ &
 $1.760 \pm 0.474$ &
 $0.530 \pm 0.325$ &
 $1.030 \pm 0.570$ &
 $4.210 \pm 3.162$
 \vspace*{0.0mm} \\
M-$p_{qpm}$ &
 $1.614 \pm 0.422$ &
 $0.408 \pm 0.294$ &
 $1.020 \pm 0.534$ &
 $5.536 \pm 3.162$
 \vspace*{0.0mm} \\
 \hline 
 \end{tabular}
 
 \end{table}

\begin{table}[ht!]
\centering
   \caption{\label{tab:Validation_test_data_mock_LowZCMASSeLRG_mocks_only}  Validation test for data below $z<0.8$. The first column shows the configuration of the measurement of cosmogical parameters ( "M" for mock catalogues) with different fiducial cosmologies ($p_F$, $p_{F2}$, $p_{qpm}$ correspond to equations \ref{fid-cosmo}, \ref{eq:fid-cosmology-test}, \ref{qpm-cosmo}, respectively). Each of the 2-4 columns shows the mode of the variable of each chain and each $98\%$ percentile divided by the fiducial cosmology of the mocks. The last column shows the $\chi^2$ test in respect of the $n.d.f$. }

\hspace{1.2cm}
\begin{tabular}{l c c c c} \hline
$\mathcal{R}_H/d_V$  & $b_0$ & $\Omega_m/\Omega_{m,qpm}$ & $\Omega_{\Lambda}/\Omega_{\Lambda,qpm}$ & $\chi^2\pm\sqrt{2.ndf}$, ndf=5 \\ 
 \hline
M-$p_F$ &
 $1.827 \pm 0.484$ &
 $1.594 \pm 1.132$ &
 $1.521 \pm 0.783$ &
 $3.731 \pm 3.162$
 \vspace*{0.0mm} \\
M-$p_{F2}$ &
 $1.760 \pm 0.474$ &
 $1.829 \pm 1.119$ &
 $1.451 \pm 0.802$ &
 $4.210 \pm 3.162$
 \vspace*{0.0mm} \\
M-$p_{qpm}$ &
 $1.614 \pm 0.422$ &
 $1.407 \pm 1.015$ &
 $1.437 \pm 0.753$ &
 $5.536 \pm 3.162$
 \vspace*{0.0mm} \\
 \hline 
 \end{tabular}
 
 \end{table}

\subsection{Definition of the Fractal Correlation Dimension}\label{sec:Definition_of_the_Fractal_Correlation_Dimension}

\citet{theiler1990estimating} and references therein, have shown the vast and unexplored by physicists regions of fractal dimension, on several flavours and definitions of fractals from the generalised fractal dimension definition to the information dimension definitions. The latter definitions capture the idea of the information entropy. In particular, he has shown that the p-point correlation functions, can be mathematically visualised by fractal p-point correlation dimensions see equation 51. In this study, we have focused in the two point statistics of a distribution of a tracers, such the one of galaxies. Two points statistics are statistics that are widely use in the literature of the large scale structures. The fractal two-point point correlation dimension, or simply expressed as fractal correlation dimension or fractal dimension, $\mathcal{D}_2(r)$, is the one which corresponds to the two-point correlation function, $\xi(r)$, which is widely used in LSS science, see for e.g. \citep{WiggleZ}. Therefore, our estimator of the fractal dimension is equivalent to the one of the fractal 2-point correlation dimension, denoted by \refEq{eq:d2observable}. In a future study, it would be interesting to study fractal correlation dimension of higher orders. It would also interesting to study other flavours of fractals, such as the one of the information fractals.

\subsection{Integral constraint bias}\label{sec:Integral_Constrain}

A note on a publication \citep{2020arXiv200615022H} which appear when this work was under review. In particular, \citet{2020arXiv200615022H} has shown that estimators of normalised number counts, which in a theoretical form is given by 
\begin{equation}\label{eq:normalised_counts_in_sphere}
	\mathcal{N}(<r)=1 + \frac{3}{r^3}\int^r_0 \xi(s)s^2 ds
\end{equation}
potentially are biased due to the integral constraint, namely integral constraint bias. This means that the amplitude of this functional changes according to a certain value, and therefore the homogeneity scale estimated from these observables is biased approximately to $40\%$. This potentially affects our estimates since the normalised counts in spheres are related to the fractal dimension, $\mathcal{D}_2(r)$, since 
\begin{equation}
	\mathcal{D}_2(r)\equiv \frac{d\ \ln}{d\ \ln r} \mathcal{N}(<r) + 3 \; .
\end{equation}
This integral constraint bias which is proposed by the authors can be introduced as a simple parameter, $b_{\rm i. c.}$ to the aforementioned model, i.e. in the \refEq{eq:normalised_counts_in_sphere} as follows. The integrand of \refEq{eq:normalised_counts_in_sphere} is composed by the two point correlation function, $\xi(s)$ scaled with $s^2$. As shown by \refEq{eq:kaiser_model}, the two-point correlation function becomes for a redshift space distorted galaxy sample:
\begin{equation}
\xi_0^{(s,G)}(s,z) = b^2(z)\left[ 1 + \frac{2}{3}\frac{f(z)}{b(z)} + \frac{1}{5}\left( \frac{f(z)}{b(z)} \right)^2  \right]\xi_0^{(r,m)}(s,z)
 \end{equation}
 In the case of a simple redshift evolved bias model, i.e. $b(z)=b_0\sqrt{1+z}$, we have:
\begin{equation}
\frac{\xi_0^{(s,G)}(s,z)}{\xi_0^{(r,m)}(s,z)} = b_0^2 (1+z) \left[ 1 + \frac{2}{3}\frac{f(z)}{b_0 \sqrt{1+z}} + \frac{1}{5}\left( \frac{f(z)}{b_0 \sqrt{1+z}} \right)^2  \right]
 \end{equation} 
 Now we introduce the $b_{\rm i.c}$ to the bias model, i.e. $b(z)=b_{\rm i.c} b_0 \sqrt{1+z}$ and we have:
 \begin{equation}\label{eq:modified_parameter_to_account_int_cont_bias}
\frac{\xi_0^{(s,G)}(s,z)}{\xi_0^{(r,m)}(s,z)} = b_{\rm i.c}^2b_0^2 (1+z) \left[ 1 + \frac{2}{3}\frac{f(z)}{b_{\rm i.c}b_0 \sqrt{1+z}} + \frac{1}{5}\left( \frac{f(z)}{b_{\rm i.c}b_0 \sqrt{1+z}} \right)^2  \right]
 \end{equation} 
 Now plugging \refEq{eq:modified_parameter_to_account_int_cont_bias} to \refEq{eq:d2observable_galaxies}, i.e. $\mathcal{D}^{(s,G)}(r)$, instead of \refEq{eq:kaiser_model} and following our methodology, we expect to estimate the quantity $b_0b_{\rm i.c.}$  instead the quantity $b_0$, using the likelihood analysis, see \refS{sec:MCMC}. Therefore, we conclude that the estimation of the cosmological parameters, $(\Omega_m,\Omega_{\Lambda})$ remain the same and the change of the amplitude due to the integral constraint is going to be absorbed by the $b_0b_{\rm i.c.}$ parameter.

A more systematic study is required to assess quantitatively this effect. Finally, this integral constraint bias does not affect the main conclusion of our study, which is the following. Our study shows the complementary of the homogeneity scale with respect to other cosmological probes.



\label{Bibliography}


\bibliographystyle{unsrtnat} 
\bibliography{Bibliography} 

\end{document}